\documentclass[twocolumn]{autart}    

\usepackage{graphicx}  
\usepackage{tikz}
\usetikzlibrary{arrows,calc}
\usepackage[switch]{lineno} 
\usepackage{enumerate} 
\usepackage{amsmath}
\usepackage{amssymb} 
\usepackage{subfig}
\usepackage{color}
\usepackage{cases}
\usepackage{comment}

\newcommand{\BL}{\color{black}}

\begin{document}

\begin{frontmatter}

\title{Limit Cycles Analysis and Control  of Evolutionary Game Dynamics with Environmental Feedback \thanksref{footnoteinfo}} 
\thanks[footnoteinfo]{The work of M. Cao is supported in part by the European Research Council (ERC-CoG-771687).
L. Gong, W. Yao, and J. Gao are funded by China Scholarship Council (CSC).}

\author[First]{Lulu Gong}, 
\author[First]{Weijia Yao}, 
\author[Second]{Jian Gao}, 
\author[First]{Ming Cao} 

\address[First]{ENTEG,  Faculty of Science and Engineering, University of Groningen, 9747 AG, Groningen, The Netherlands, (e-mail: lukegong20@gmail.com, weijia.yao.new@outlook.com, and m.cao@rug.nl).}
\address[Second]{School of Science, Beijing University of Posts and Telecommunications, Beijing 100876, China, (e-mail: gao.jian@bupt.edu.cn).}


\begin{abstract}                
Recently, an evolutionary game dynamics model taking into account the environmental feedback has been
proposed to describe the co-evolution of strategic  actions of a population of individuals and the state of  the surrounding environment;  correspondingly a range of interesting dynamic behaviors  have been reported. In this paper, we provide new theoretical insight into such behaviors and  discuss control options.
Instead of the standard  replicator dynamics, we use a  more realistic and comprehensive model of \emph{replicator-mutator dynamics}, to describe the strategic evolution of the population. After integrating the environment feedback, we study the effect of mutations on the resulting closed-loop system dynamics. We prove the conditions for two types of bifurcations, \emph{Hopf bifurcation} and \emph{Heteroclinic bifurcation},  both of which result in stable \emph{limit cycles}. These limit cycles have not been  identified in existing works, and we  further prove that  such limit cycles are in fact  persistent in a large parameter space and are almost globally stable.  In the end, an intuitive control policy based on incentives is applied, and the effectiveness of this  control policy is  examined  by analysis and simulations. 
\end{abstract}


\end{frontmatter}

\section{Introduction}
Evolutionary game theory studies the  evolution of strategic decision-making processes of individuals in one or multiple populations. A range of mathematical models have been  extensively investigated, among which the \emph{replicator dynamics} model is the most well-known \cite{Hofbauer:98}. Powerful theories and tools originating from evolutionary game theory facilitate the study of complex system behaviors of biological, ecological, social, and engineering fields \cite{Nowak:01}, \cite{Komarova:04}, \cite{Lee:19},  \cite{Pais:12},  \cite{stella2018bio}.
In the classic game setting, the payoffs in each  two-player game are usually predetermined in the form of constant payoff matrices. However, in many applications, especially  those involving common resources  in the environment that are consumed by groups of individuals, it is recognized that the payoffs can change over time or be affected directly by  external environmental factors. Thus game-playing individuals' decisions can influence the surrounding environment, and the environment,  especially its richness in resources, also acts back on the payoff distributions, which forms the so-called bi-directional \emph{game-environment feedback}   \cite{Weitz:16}. Such a feedback mechanism is believed to be able to explain rich real-world  complex population dynamics, including human decision-making, social culture evolution, plant nutrient acquisition, and natural resource harvesting.
In particular, evolutionary dynamics with game-environment feedback  have attracted great attention in recent years,  due to its high relevance in biological and sociological systems
\cite{rand2017}, \cite{Lee:19}, \cite{Tilman:20}. 

This line of research  is originated in  \cite{Weitz:16}, where the replicator dynamics of a two-player two-strategy game are coupled with the logistic environment resource dynamics.  Utilizing an elaborate environment-dependent payoff matrix,  the joint game-environment dynamics exhibit interesting system behaviors. In most cases, the strategic states of the population and the environment converge to an equilibrium point on the boundary  of the phase space with  the zero value of environment state, which reflects the \emph{tragedy of the commons}. The system dynamics may also show cyclic oscillations which correspond to closed
periodic orbits. Particularly when the system dynamics  converge to a heteroclinic cycle on the boundary, the environment state  cycles between low and high  values, which is referred to as the \emph{oscillating tragedy of commons}.  Since then, \cite{Weitz:16} has been followed by several extensions and variations \cite{Hauert:19}, \cite{Tilman:20}, \cite{Gong:18},  \cite{Kawano:19},  \cite{Muratore:19}, \cite{Lin2019}.  The game-environment feedback mechanism has been studied more broadly in different applications,  while its theoretical settings have  been  adapted in different research directions, e.g. implementing different types of environment resource models and considering the structures of the populations. New meaningful dynamical behaviors  have been identified, and these results are in turn helpful to understand real-world applications.

The majority of the existing studies on the game-environment framework chooses replicator dynamics to model the strategic dynamics of the game-environment system.
While the replicator dynamics have been proved to be a powerful model in analyzing a variety of classical games, they do not take into account the mutation or exploration in strategies which exists ubiquitously in biological and sociological systems.
Genetic mutations typically occur with small  but non-negligible probabilities, and random exploration of available strategies  is common in behavioral experiments \cite{Traulsen-09}. Mutations and explorations can be captured by allowing individuals to spontaneously switch from one strategy to another in small probability, which results in the so-called \emph{replicator-mutator dynamics}. It has also played a prominent role in evolutionary game theory and  has been applied frequently in different fields   \cite{Nowak:01},  \cite{Komarova:04}, \cite{Pais2011}, \cite{Pais:12}.

We  note that although rich system behaviors, such as convergence to equilibria or heteroclinic cycles on the boundary and existence of neutral periodic orbits, have been identified in related works, the results on the limit cycle dynamics have been rarely reported in the game-environment feedback setting. 
Only in \cite{Tilman:20}, it is shown that the time-scale difference between game and environment dynamics can result in limit cycles. When  mutations are taken into account, it is of great interest to study if the coupled game and environment dynamics can exhibit limit cycles.

 We also emphasize that research on how to design reasonable and effective control policies to achieve expected system behaviors in evolutionary games has  attracted increasing attention \cite{Morimoto2016}, \cite{ZHU201694}, \cite{Riehl2016},  \cite{Riehl2019}.
 Specifically, using game-environment feedback,  \cite{Paarporn:18} considers  optimal control problems where  control takes the form of the incentives and opinions respectively. The adopted control policies can  maximize  accumulated resource over time. However, since the controlled system  still exhibits  heteroclinic oscillations, repeated collapses of the resource are inevitable.
\cite{Wang:20} considers  adjusting the law of feedback from population states to the environment based on a slightly different co-evolutionary model of public goods games.  An involved nonlinear control law is proved to be able to steer  the system to evolve towards the designed  behaviors, but the biological implication of such control policies still remains open to debate.

Motivated by all the  unsolved issues just mentioned, in this work we study the simultaneous co-evolution of game and environment  using an integrated model
consisting of replicator-mutator dynamics and logistic resource dynamics. Although similar linear environment-dependent payoff matrices have been  studied before,  we focus on investigating the significant impact of mutations on the overall system dynamics using bifurcation theory as the main tool. In particular, we show that two different types of bifurcations, which lead to limit cycles, can take place.  In sharp comparison to the marginal oscillating behaviors reported before, such as neutral periodic orbits and heteroclinic cycles, we clarify that  the identified limit cycle conditions correspond to a large space of the mutation parameter and such limit cycles are almost globally stable. In this sense, the limit cycle behavior is robust and persistent,   which agrees with biological observations. Furthermore, when mutations are not too rare, the corresponding limit cycle will not be close to the boundary,  and consequently the (oscillating) tragedy of commons can be averted. 

 Enabled by the new insight into co-evolutionary dynamics of the game and environment,  we consider further a simple  but effective incentive-based control policy, where the individuals receive some external incentives in the game interactions if they choose the cooperation strategy.  It is shown that the controlled system dynamics can converge to certain equilibria with a high level of the environmental resource.  Thus, the effectiveness of the proposed control policy is guaranteed. Compared with the proposed control policies in [\cite{Paarporn:18}] and [\cite{Wang:20}], our proposed approach is not only more effective, but also
 easier to implement than influencing the opinions of the individuals or realizing more complicated feedback between strategists and the environment.

Our preliminary conference paper  \cite{Gong:2020} considers Hopf bifurcation and limit cycles of the system dynamics in a special case where the enhancement effect and the degradation effect  is balanced, but in this paper we  consider general bifurcation conditions
and provide a comprehensive account of the results. Moreover, the latter part about incentive-based control of this paper is  not in the conference version.

The remainder of this paper is organized as follows. Section $2$ provides the framework of the co-evolutionary dynamics, and introduces the integrated model with game-environment feedback. The analysis of system dynamics is presented in Section $3$. The incentive-based  control is applied  and its performance is provided in Section $4$. Finally, concluding remarks and discussion points are given in Section $5$.

\section{Problem Formulation}
\begin{figure}[htbp!]
    \centering
   	\includegraphics[width=5cm]{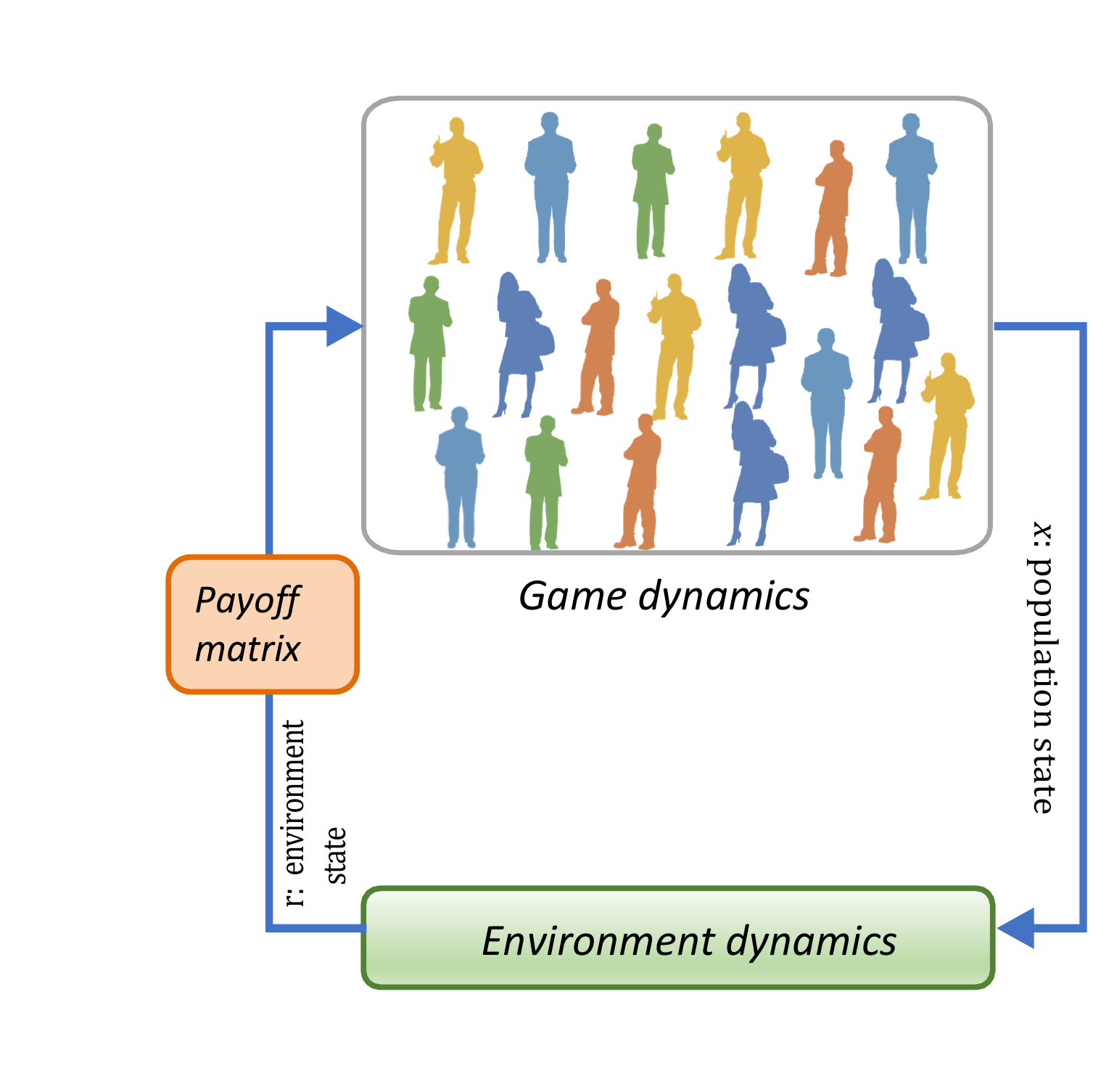}
    \caption{The framework of evolutionary game dynamics with the environmental feedback}
    \label{fig:framework}
\end{figure}

Co-evolutionary game and environment dynamics occur because evolutionary game dynamics are sometimes coupled with the dynamics of the surrounding environment (see Fig. \ref{fig:framework} for an illustrative drawing). 
The coupled game and environment dynamics  form a closed-loop system,  which  consists of  bi-directional actions between the environment and the  population's game play: an individual's fitness depends on not only the population state, but also  the state of the environment, and the state of the environment is influenced by the  distribution  of the choices of strategies in the population.
 In the following subsections, we will introduce each component of this system in turn,  and then integrate them to give a complete description of our model.

\subsection{Environment-dependent payoffs}
We consider games played in a changing environment,  which is characterized by the richness $r\in [0,1]$ of a resource of interest; such a resource  has direct influence on the payoffs that players receive. Correspondingly,  the payoffs  become dynamic and depend on $r$. We consider a two-player game with two strategies, \emph{Cooperation} and \emph{Defection} (C, D).
Following \cite{Weitz:16}, we assume that the environment-dependent payoff matrix is   a convex combination of the payoff matrices from the classic Prisoner's Dilemma, i.e., 
\begin{equation} \label{eq:matrix}
\begin{aligned}
A(r)&=(1-r)\begin{bmatrix}
R_1&S_1\\
T_1&P_1
\end{bmatrix}+r\begin{bmatrix}
R_2&S_2\\
T_2&P_2
\end{bmatrix}\\
&=\begin{bmatrix}
R_1+(R_2-R_1)r&S_1+(S_2-S_1)r\\
T_1+(T_2-T_1)r&P_1+(P_2-P_1)r
\end{bmatrix},
\end{aligned}
\end{equation}
where $R_1>T_1$, $S_1>P_1$, $R_2<T_2$ and $S_2<P_2$.   The entries of $A(r)$ represent the payoffs to the players at the given environment state $r$. For example, a C player receives the payoff of $R_1+(R_2-R_1)r$ and $S_1+(S_2-S_1)r$ when the opponent uses strategies C and D, respectively. Because of the inequalities below (\ref{eq:matrix}), the two matrices $\begin{bmatrix}
R_2&S_2\\
T_2&P_2
\end{bmatrix}$ and $\begin{bmatrix}
R_1&S_1\\
T_1&P_1
\end{bmatrix}$  correspond to the prisoner's dilemma and its inverse game, respectively.  
Note that the payoff matrix $A(r)$ is the linear interpolation between the above two matrices. Thus, when $r$ approaches $0$ and $1$, the mutual cooperation and defection tend to be the \emph{Nash Equilibrium} of this dynamic game respectively.  The biological interpretation is that: the individuals tend to cooperate  facing the depleted environmental resource, and  become more selfish and choose to defect while the resource becomes abundant.  
\subsection{Replicator-Mutator equations}
The two-player game governed by the payoff matrix $A(r)$ in (\ref{eq:matrix}) are played in a well-mixed and infinite population.  We denote the proportion of individuals choosing C  by the variable $x\in [0,1]$. Since there are only two strategies, the population state can be represented by the vector $\mathbf{x}=[x~~ 1-x]^\top$.  Assume the individuals with different strategies can mutate into each other with an identical probability $\mu \in [0,1]$: a fraction of cooperators, $\mu x$, spontaneously change to choose strategy D; a fraction of defectors, $\mu (1-x)$, change to choose strategy C at the same time.
We focus on the dynamics  of $x$, which can be  described by the  following replicator-mutator equation
\begin{equation}\label{eqn:rd}
\dot{x}=x[(A(r)\mathbf{x})_{1}-\mathbf{x}^{\top}A(r)\mathbf{x}]-\mu x+\mu(1-x), 
\end{equation}
where $(A(r)\mathbf{x})_{1}$ is the first entry of $A(r)\mathbf{x}$ and represents the fitness of choosing strategy C,  and  $\mathbf{x}^{\top}A(r)\mathbf{x} $ is the average fitness at the population state $\mathbf{x}$.  The term $-\mu x$ represents the mutation from strategy C towards D, while $\mu(1-x)$ captures the  mutation in the opposite direction, from D towards C.

\subsection{Closed-loop system with game-environment feedback}
To model the evolution of the environmental change, we use the standard logistic model 
\begin{equation}\label{eqn:envi}
\dot{r}=r(1-r)[\theta x-(1-x)],
\end{equation}
where the parameter $\theta>0$ represents the ratio between the enhancement effect due to cooperation and degradation effect due to defection.  The influence of the population on the environment state is captured by the function $[\theta x-(1-x)]$, and the dynamics are restricted in the unit interval by $r(1-r)$. 
When $\theta=1$, the enhancement effect and the degradation effect are balanced. And if $\theta>1$ or $0<\theta<1$,  the enhancement effect is respectively stronger or weaker than the opposite.   

Substituting  (\ref{eq:matrix}) into (\ref{eqn:rd}), and then combining  with (\ref{eqn:envi}), we obtain the closed-loop planar system describing the evolutionary game dynamics under the game-environment feedback: 
\begin{equation}\label{eqn:closeloop}
\begin{cases}
&\begin{aligned}
   \dot{x}=&x(1-x)[xr(-c+d-a+b)+x(a-b)\\
   & -r(d+b)+b]+\mu(1-2x)
\end{aligned}\\
&\dot{r}=r(1-r)[\theta x-(1-x)],
\end{cases}
\end{equation}
where to simplify notations, we have used $a=R_1-T_1$, $b=S_1-P_1$, $c=T_2-R_2$, and $d=P_2-S_2$,  all of which are positive because of the inequalities below (\ref{eq:matrix}). Note that system (\ref{eqn:closeloop}) reduces to the model considered in \cite{Weitz:16} when $\mu=0$. Therefore, system (\ref{eqn:closeloop}) generalizes the previous model.

 In view of the payoff matrix (\ref{eq:matrix}) and system equations (\ref{eqn:closeloop}), these four parameters for re-parameterization have intuitive interpretations with respect to  strategy changes  when the resource is either abundant or depleted: the parameter $a$ quantifies the incentive to stick to strategy C given
that all individuals are following strategy C and the system is currently in a resource-poor state; $b$ quantifies the incentive to switch to strategy C when  the resource is depleted and all individuals enforce strategy D; in contrast, $c$ quantifies the incentive to switch to strategy D given that all individuals enforce strategy C and the system is in a state of sufficient resource; $d$ quantifies the incentive to stick to strategy D  when the resource is abundant and all individuals are following strategy D.

\section{System Dynamics and Bifurcations}
The state space of  system (\ref{eqn:closeloop}) is the unit square $\mathcal{I}= [0, 1]^2$.
The interior region of this square is denoted by $\text{int}(\mathcal{I})=(0,1)^2$.
Four sides of the unit square form the boundary $\partial \mathcal{I}$. 
For convenience of exposition, we denote the four sides by: 
\begin{equation}\label{4sides}
    \begin{aligned}
    &\mathcal{B}_t=\{(x,r):x\in[0,1],r=1\},\\
    &\mathcal{B}_b=\{(x,r):x\in[0,1],r=0\},\\
    &\mathcal{B}_l=\{(x,r):x=0, r\in[0,1]\},\\
    &\mathcal{B}_r=\{(x,r):x=1, r\in[0,1]\}.
    \end{aligned}
\end{equation}
Then, one has $\partial \mathcal{I}:= \mathcal{B}_t \cup \mathcal{B}_b\cup \mathcal{B}_l \cup \mathcal{B}_r$.
It is easy to check that $\mathcal{I}$
is invariant under the dynamics (\ref{eqn:closeloop}) for all $\mu \in [0,1]$ by Nagumo's theorem \cite[Theorem 4.7]{blanchini:08}.

System (\ref{eqn:closeloop}) can have equilibria  in the interior $\text{int}(\mathcal{I})$;  we call them  \emph{interior equilibria}. 
System (\ref{eqn:closeloop})  also has equilibria on the boundary $\partial\mathcal{I}$ (\emph{boundary equilibria}), which will be discussed in detail in Section 3.4.
 It is easy to check that the interior equilibria,  if they exist, must lie on the line $\{(x,r)\in \text{int}(\mathcal{I}): x=1/(\theta+1)\}$. Substituting such $x$ into the {\BL $x$-dynamics in (\ref{eqn:closeloop}), we have}
\begin{equation*}
    \begin{aligned}
   0=&x(1-x)[xr(-c+d-a+b)\\
   & +x(a-b)-r(d+b)+b]+\mu(1-2x),
\end{aligned}
\end{equation*}
 which leads to
\[r=\frac{\theta a+\theta^2 b+\mu(\theta^3+\theta^2-\theta-1)}{\theta(a+c+\theta b+\theta d)}.\] 
Thus if the following condition holds
\begin{equation}\label{condition1}
-(\theta a+\theta^2 b)<\mu(\theta^3+\theta^2-\theta-1)<\theta c+\theta^2 d,
\end{equation}
system (\ref{eqn:closeloop}) has a unique interior equilibrium
\begin{equation}\label{interiorequilibrium}
(x^*,r^*)=\left(\frac{1}{\theta+1}, \frac{\theta a+\theta^2 b+\mu(\theta^3+\theta^2-\theta-1)}{\theta(a+c+\theta b+\theta d)} \right).  
\end{equation}
Otherwise, no interior equilibrium exists.   The interior equilibrium supports the coexistence of the two strategies under a non-depleted resource state. In view of (\ref{condition1}), one can see that for a given $\theta$, the existence of the interior equilibrium is determined by not only the payoff parameters but also the mutation rate.

Note that when $\theta=1$, $(x^*,r^*)$ is independent of the parameter $\mu$, and (\ref{condition1}) holds trivially for all $\mu\in[0,1]$. This specific case has been studied in our previous conference paper \cite{Gong:2020}. In this paper we do not require $\theta=1$ and study the generic $\theta$ that takes any positive value satisfying (\ref{condition1}).

Towards the goal of establishing conditions for the existence of limit cycles, we  first present some preliminary results on the non-existence of closed orbits or limit cycles.
\subsection{Non-existence of closed orbits}
First, we define the $\mathbf{C}^1$ function $\varphi(x,r): \text{int}(\mathcal{I})\rightarrow \mathbb{R}$ by
\begin{equation}\label{dulacfunction1}
    \varphi(x,r)=x^{\alpha} (1-x)^{\beta} r^{\gamma} (1-r)^{\delta}
\end{equation}
with the parameters $\alpha$, $\beta$, $\gamma$, and $\delta$ to be determined later. Note that $\varphi(x,r)$ is strictly positive in $\text{int}(\mathcal{I})$ for any values of these parameters.
Denote the right hand side of (\ref{eqn:closeloop}) by $f(x,r)=[f_1(x,r),f_2(x,r)]^\top$, and then the divergence of the modified vector field $\varphi f(x,r)$ on $\text{int} (\mathcal{I})$ is given by 
\begin{equation}\label{divergence}\begin{aligned}
&\text{div}~\varphi  f(x,r)=\frac{\partial (\varphi f_1)}{\partial x}(x,r)+\frac{\partial (\varphi f_2)}{\partial r}(x,r)\\
&=\varphi(x,r)(-(\alpha+\beta+3)(-c+d-a+b)x^2r\\
&~~~-(\alpha+\beta+3)(a-b)x^2\\
&~~~+[(-c+2d-a+2b)\alpha+(d+b)\beta \\
&~~~-(\theta+1)(\gamma+\delta+2)+2(-c+2d-a+2b)]xr\\
&~~~+[(a-2b)\alpha-b\beta+(\theta+1)(1+\gamma)+2a-4b]x\\
&~~~+[-(d+b)\alpha+\gamma+\delta+2-b-d]r\\
&~~~+\frac{\mu(1-2x)(\alpha-\alpha x-\beta x)}{x(1-x)}\\
&~~~+(1+\alpha) b-(1+\gamma)-2\mu).
\end{aligned}
\end{equation}
Let all the coefficients of the  powers of $x$, $r$ in (\ref{divergence}) (except for the term containing $\mu$) be zero, and we obtain the following set of equations
\begin{equation}\label{eqn:linearequations}
    \begin{aligned}
    (\alpha+\beta+3)(-c+d-a+b)&=0\\
    (\alpha+\beta+3)(a-b)&=0\\
    (-c+2d-a+2b)\alpha+(d+b)\beta \\
-(\theta+1)(\gamma+\delta+2)+2(-c+2d-a+2b)&=0\\
   (a-2b)\alpha-b\beta+(\theta+1)(\gamma+1)+2a-4b&=0\\
   -(d+b)\alpha+\gamma+\delta+2-b-d&=0.
    \end{aligned}
\end{equation}
We first show that the equation set (\ref{eqn:linearequations}) always has solutions.
\begin{lem}\label{alphabetagammadelta}
There always exist some $\alpha$, $\beta$, $\gamma$, and $\delta$ as solutions to (\ref{eqn:linearequations}).
\end{lem}
\begin{pf}
See Appendix A.
\hfill$\qed$
\end{pf}

In fact, if one of $-c+d-a+b$ and $a-b$ is not zero, (\ref{eqn:linearequations})  will have exactly one solution
\begin{equation}\label{alphavalue}
    \begin{aligned}
    &\alpha=-\frac{\zeta+c+a}{\zeta},\\
    &\beta=\frac{-2\zeta+c+a}{\zeta},\\
    &\gamma=\frac{-(a+\theta+1)\zeta+a^2+ac-ab-bc}{(\theta+1)\zeta},\\
    &\delta=-\frac{(a+c)(\theta b+\theta d+d+a)+(\theta +1-a)\zeta}{(\theta+1)\zeta}
    \end{aligned}
\end{equation}
with $\zeta=a+c+\theta b+\theta d>0$. Otherwise,  (\ref{eqn:linearequations}) will have infinitely many solutions including (\ref{alphavalue}). So we fix the parameters $\alpha$, $\beta$, $\gamma$, and $\delta$ at the values given in (\ref{alphavalue}) such that the divergence (\ref{divergence}) can be greatly simplified. 

Then we show the non-existence of closed orbits for system (\ref{eqn:closeloop}).
\begin{lem}\label{nonclosedorbit}
For system (\ref{eqn:closeloop}), the following statements hold:
\begin{enumerate}
    \item when $ad-bc>0$,
there exists some $\mu_0>0$ such that 
there are no closed orbits in $\mathcal{I}$ for $\mu \in [\mu_0,1]$;
\item when $ad-bc<0$,
there are no closed orbits in $\mathcal{I}$ for $\mu\in [0,1]$;
\item when $ad-bc=0$,
there are no closed orbits in $\mathcal{I}$ for $\mu\in (0,1]$.
\end{enumerate}
\end{lem}
\begin{pf}
The proof is direct by applying  Bendixson-Dulac criterion \cite[Theorem 4.1.2]{Wiggins:00}. Since the boundary $\partial \mathcal{I}$ contains equilibria (as shown in Section 3.4), it is not possible to have periodic orbits on it. And the sides are either invariant or repelling, so they cannot intersect with  closed orbits. Now let us consider the interior $\text{int}(\mathcal{I})$ which is simply connected.

 In view of the fact that $\alpha$, $\beta$, $\gamma$, and $\delta$ are given as in (\ref{alphavalue}), one can calculate the divergence of $\varphi f(x,r)$ 
\begin{equation}\label{divergence1}
\begin{aligned}
& \text{div}~\varphi  f(x,r)\\
&=\varphi(x,r)\bigg( \frac{\mu(1-2x)(3x+\alpha)}{x(1-x)}+(1+\alpha) b \\
&\quad \quad \quad \quad \quad-(1+\gamma)-2\mu   \bigg)\\
&=\varphi(x,r)\left(\frac{\mu(1-2x)(3x+\alpha)}{x(1-x)}-2\mu-\frac{\theta(bc-ad)}{(\theta+1)\zeta}\right).
 \end{aligned}
\end{equation}
From (\ref{alphavalue}) one knows $\alpha\in(-2,-1)$, and thus the term $g(x):=\frac{\mu(1-2x)(3x+\alpha)}{x(1-x)}$  takes its maximum value at $\bar{x}\in(0,1)$  given by
\begin{equation}\label{xbar}
    \bar{x}=\begin{cases}
    &\frac{\alpha- \sqrt{-\alpha(3+\alpha)}}{3+2\alpha}, ~-1.5<\alpha<-1,\\
    &0.5, ~\quad\quad\quad\quad\quad\quad~~~ \alpha=-1.5,\\
    &\frac{\alpha+ \sqrt{-\alpha(3+\alpha)}}{3+2\alpha}, ~-2<\alpha<-1.5.
    \end{cases}
\end{equation}
Rearranging (\ref{divergence1}) yields 
\begin{equation}\label{divergence2}
\begin{aligned}
 \frac{\text{div}~\varphi  f(x,r)}{\varphi(x,r)}&\leq\frac{\mu(1-2\bar{x})(3\bar{x}+\alpha)}{\bar{x}(1-\bar{x})}-2\mu-\frac{\theta(bc-ad)}{(\theta+1)\zeta}\\
 &=\frac{(1-2\alpha)\bar{x}+\alpha-4\bar{x}^2}{\bar{x}(1-\bar{x})}\mu -\frac{\theta(bc-ad)}{(\theta+1)\zeta}.
\end{aligned}
\end{equation}
It is easy to check that $\frac{(1-2\alpha)\bar{x}+\alpha-4\bar{x}^2}{\bar{x}(1-\bar{x})}$ is negative for any value of $\alpha \in(-2,-1)$.
Since $\varphi(x,r)$ is positive in $\text{int}(\mathcal{I})$ by definition, if $ad-bc>0$, 
from (\ref{divergence2}) one has $\text{div}~\varphi f(x,r)\leq 0$ when 
\begin{equation}
\mu\geq\mu_0:= \frac{\theta(bc-ad)\bar{x}(1-\bar{x})}{(\theta+1)\zeta[(1-2\alpha)\bar{x}+\alpha-4\bar{x}^2]}.   
\end{equation} And because the divergence is not identically zero in $\text{int}(\mathcal{I})$, from the Bendixson-Dulac criterion,  one  knows that no closed orbits can exist in $\text{int}(\mathcal{I})$ for $\mu\geq\mu_0$.

For the second case, it is easy to prove since the right hand side of (\ref{divergence2}) will  always be negative for any $\mu\geq 0$.  For the third case, the right hand side of (\ref{divergence2}) is always negative when $\mu>0$.
\hfill$\qed$
\end{pf}
 In Lemma \ref{nonclosedorbit}, 
it is noted that the conditions for the non-existence of periodic orbits depend on the relationship between $ad$ and $bc$. The reason is as follows: when $ad-bc<0$, the divergence $\text{div}~\varphi  f(x,r)$ will remain positive for any $\mu\in [0,1]$ in view of (\ref{divergence2}), which excludes the possibility of periodic orbits in the phase space; when $ad-bc>0$ or  $ad-bc=0$, there exists some $\mu$ such that $\text{div}~\varphi  f(x,r)$ equals $0$. As a consequence, it is possible for the system to have periodic orbits in these two cases.  Another point worth noting is that
the only difference between the second and third cases of Lemma \ref{nonclosedorbit} is that system (\ref{eqn:closeloop})
may have periodic solutions for $\mu=0$ when $ad-bc=0$. This possibility has been shown in \cite{Weitz:16}, \cite{Gong:18}. Despite the possibility of closed orbits under certain conditions,
 for $\mu=0$ it can be proven that limit cycles (isolated closed orbits) are impossible to exist in (\ref{eqn:closeloop}) for all the parameter space as summarized below in Lemma \ref{nolimitcycle}.
\begin{lem}\label{nolimitcycle}
System (\ref{eqn:closeloop}) has no limit cycles in $\mathcal{I}$ when $\mu=0$.
\end{lem}
The full proof is presented in \cite{Gong:2020}. See Appendix \ref{proofoflemma3} for a sketch of the proof.

\subsection{Hopf bifurcation at the interior equilibrium}
In this section we take $\mu$ as the bifurcation parameter to study the effect of different mutation rates on the system dynamics. We  investigate the possible bifurcations of the dynamics (\ref{eqn:closeloop}) as $\mu$ varies in $[0,1]$. We analyze the stability of the interior equilibrium and prove that it involves a Hopf bifurcation which leads to periodic orbits.

We first invoke the Hopf bifurcation theorem \cite{Guckenheimer:00} which will be used to prove the existence of stable limit cycles.
\begin{thm}[Hopf bifurcation theorem] \label{Hopftheorem}
Suppose that the system $\dot{y}=F(y,\rho), y\in \mathbb{R}^2, \rho\in \mathbb{R}$ has an equilibrium $(y_0,\rho_0)$ at which the following properties are satisfied:
\begin{enumerate}
    \item the Jacobian $D_y F|_{(y_0,\rho_0)}$ has a simple pair of pure imaginary eigenvalues $\lambda(\rho_0)$ and $\bar{\lambda}(\rho_0)$;
    \item $\frac{d\Re(\lambda(\rho))}{d \rho}|_{\rho_0}\neq 0$.
\end{enumerate}
Then the dynamics  undergo a Hopf bifurcation at $(y_0,\rho_0)$, which results in a family of periodic solutions in a sufficiently small neighborhood of $(y_0,\rho_0)$.
\end{thm}

Then we make an assumption about some parameters of system (\ref{eqn:closeloop}). \begin{assum}\label{assumptionofallparameters}
We assume  that $
    0< ad-bc\leq \frac{\Delta}{\theta^2}.
$
\end{assum}
The following theorem demonstrates that the interior equilibrium $(x^*,r^*)$ of system (\ref{eqn:closeloop}) undergoes a Hopf bifurcation.
\begin{thm}\label{thm:hopfbifurcation}
Under Assumption \ref{assumptionofallparameters},
  the interior equilibrium $(x^*,r^*)$ of system (\ref{eqn:closeloop}) with the bifurcation parameter $\mu$ undergoes a \emph{supercritical Hopf bifurcation} at $\mu=\mu_1:=\frac{\theta^2(ad-bc)}{\Delta}$, which leads to the existence of  stable limit cycles for  $\mu<\mu_1$ in the vicinity of $\mu_1$.
\end{thm}
\begin{pf}
To prove the existence of Hopf bifurcation, we need to show that the two conditions of Theorem \ref{Hopftheorem} are satisfied in system (\ref{eqn:closeloop}). 
The Jacobian of the vector field of (\ref{eqn:closeloop}) at $(x^*,r^*)$ is given by
\begin{equation}\label{interiorJacobian1}
 J^*=\begin{bmatrix}
\frac{\theta^2(ad-bc)-\mu \Delta}{\theta(\theta+1)\zeta}&\frac{-\theta\zeta}{(\theta+1)^3}\\
\frac{(\theta+1)(c\theta+d\theta^2-\mu\hat{\theta})(a\theta+b\theta^2+\mu\hat{\theta})}{\theta^2\zeta^2}&0
\end{bmatrix},   
\end{equation}
where  \[\hat{\theta}=(\theta^3+\theta^2-\theta-1)\] 
and \[\Delta=(a+c)(1+\theta^2+2\theta^3)+(b+d)(2\theta+\theta^2+\theta^4).\]
Note that the off-diagonal entries 
\[J_{12}=\frac{-\theta\zeta}{(\theta+1)^3}<0\] 
and 
\[J_{21}=\frac{(\theta+1)(c\theta+d\theta^2-\mu_1\hat{\theta})(a\theta+b\theta^2+\mu_1\hat{\theta})}{\theta^2\zeta^2}>0\] 
because of (\ref{condition1}).
One can calculate the eigenvalues of $J^*$ 
\begin{equation}\label{eigenvalues}
\begin{aligned}
    \lambda(\mu)&=\frac{J_{11}\pm \sqrt{J_{11}^2+4J_{12}J_{21}}}{2}\\
    &=\Re(\lambda(\mu))\pm \Im(\lambda(\mu))i.
\end{aligned}
\end{equation}
It is easy to check that when $\mu=\mu_1:=\frac{\theta^2(ad-bc)}{\Delta}$ the eigenvalues are purely imaginary, i.e.,
\begin{equation*}
    \begin{aligned}
    &\Re(\lambda(\mu_1))=0,\\
    &\Im(\lambda(\mu_1))=\sqrt{\frac{(c\theta+d\theta^2-\mu\hat{\theta})(a\theta+b\theta^2+\mu\hat{\theta})}{\theta(\theta+1)^2\zeta}},
    \end{aligned}
\end{equation*} 
which implies that the first condition listed in Theorem \ref{Hopftheorem} is satisfied.
In addition,  the second condition  is also satisfied because
\[\left.\frac{d\Re(\lambda(\mu))}{d \mu}\right|_{\mu_1}=\frac{-\Delta}{\theta(\theta+1)\zeta}< 0.\] 
 Then one can conclude that the dynamics (\ref{eqn:closeloop}) undergo a Hopf bifurcation at $\mu=\mu_1$.  Since $\mu$ varies in the interval $(0,1]$, to guarantee the occurrence of Hopf bifurcation, the critical parameter value $\mu_1$ should also be constrained in this interval, i.e., $0<\mu_1\leq 1$, which leads to 
 \[0< ad-bc \leq \frac{\Delta}{\theta^2},\]
 which is exactly Assumption \ref{assumptionofallparameters}.
 The Hopf bifurcation theorem implies that a family of periodic orbits bifurcate from $(x^*,r^*)$ for some $\mu$ in the vicinity of $\mu_1$. 
Whether the bifurcated periodic orbits are stable or not is determined by the so-called \emph{first Lyapunov coefficient} $\ell_1$ at $(x^*,r^*)$ when $\mu=\mu_1$. If $\ell_1<0$, then these periodic orbits are stable limit
cycles; if $\ell_1>0$, the periodic orbits are repelling.
Following the computation procedure in Appendix \ref{Lyapunovcoefficient}, we obtain  $\ell_1(\mu_1)$ as below
\begin{equation}\label{firstlyc}
\begin{aligned}
  \ell_1(\mu_1)&=\\
  &\frac{\theta(bc-ad)\zeta((b+d)(\theta^3+2\theta)+(a+c)(2\theta^3+1))}{2\omega_0^3(\theta+1)^4\Delta},
  \end{aligned}
\end{equation}
which is negative  because of (\ref{condition1}) and Assumption \ref{assumptionofallparameters}. 
Thus, the  bifurcated periodic orbits are  stable limit cycles.

Furthermore, in view of (\ref{eigenvalues}),  for $\mu<\mu_1$, $(x^*,r^*)$ is unstable since the eigenvalues have positive real parts. On the other hand, the interior equilibrium  is asymptotically stable for $\mu>\mu_1$ with the negative eigenvalues. Hence, the stable limit cycles exist for $\mu<\mu_1$ in the vicinity of $\mu_1$. Since the bifurcation is associated with  stable limit cycles, it is a \emph{supercritical Hopf bifurcation}.
\hfill$\qed$
\end{pf}
 The relationship between $ad$ and $bc$ has played an important role in Theorem \ref{thm:hopfbifurcation} in the sense that Hopf bifurcation can only happen when $ad-bc>0$. In view of the Jacobian (\ref{interiorJacobian1}), when $ad-bc \leq 0$, $J_{11}$ will remain non-positive no matter what value $\mu$ takes. Thus, as $\mu$ varies in $[0,1]$, the stability of the interior equilibrium does not change, which excludes the existence of a Hopf bifurcation. On the other hand, when $ad-bc>0$, to ensure there is a Hopf bifurcation for $\mu \in [0,1]$, Assumption \ref{assumptionofallparameters} has to be satisfied as proved.

From the Hopf bifurcation theory, we know that the  limit cycles generated from Hopf bifurcation have the amplitude of  $O(\sqrt{\mu_1-\mu})$. Since Hopf bifurcation is a local bifurcation, the obtained results only hold for $\mu$ in the vicinity of $\mu_1$. Next, we will show that  system (\ref{eqn:closeloop}) exhibits another type of bifurcation.

\subsection{Heteroclinic bifurcation}
In the case $\mu=0$, it has been shown  in Lemma \ref{nolimitcycle} that the system (\ref{eqn:closeloop}) cannot exhibit limit cycles for any parameters. We show in this section that limit cycles may occur for $\mu>0$ in the vicinity of $0$.

When $\mu=0$, system (\ref{eqn:closeloop}) has four equilibria on the boundary $\partial \mathcal{I}$. 
 Denote the four equilibria respectively by $E_1=(0,0)$, $E_2=(1,0)$, $E_3=(1,1)$, and $E_4=(0,1)$. Then, one can calculate the Jacobians at these points, which are 
 \begin{equation}
     \begin{aligned}
     &J_1=\begin{bmatrix}
b&0\\
0&-1
\end{bmatrix},&J_2=\begin{bmatrix}
-a&0\\
0&\theta
\end{bmatrix},\\
&J_3=\begin{bmatrix}
c&0\\
0&-\theta
\end{bmatrix},&J_4=\begin{bmatrix}
-d&0\\
0&1
\end{bmatrix}.
     \end{aligned}
 \end{equation}
It is easy to see that the eigenvalues of these equilibria are indeed the diagonal entries of the associated Jacobian matrices, and thus all of these equilibria are saddle points. 
Next we note that the trace of each Jacobian, which is denoted by $T_i$, is 
\[
T_1=b-1, ~T_2=\theta-a,~ T_3=c-\theta, ~ T_4=1-d.\]
A  heteroclinic cycle on the boundary denoted by $\Lambda$,  which connects the four equilibria  and has the orientation $E_1 \rightarrow E_2 \rightarrow E_3 \rightarrow E_4 \rightarrow E_1$, can be identified.  Its stability is determined by the payoff parameters.

\begin{lem}\label{heterocliniccycle}
Consider system (\ref{eqn:closeloop}) at $\mu=0$. There exists a heteroclinic cycle $\Lambda$, which is stable (resp. unstable) when the condition $ad>bc$ (resp. $ad<bc$) is satisfied.
\end{lem}
We refer  to the  reference \cite{Weitz:16} for the complete proof for the case of the stable heteroclinic cycle. And the opposite case can be proved using the analogous method according to \cite[Corollary 2]{Hofbauer1994}.

When $\mu>0$, the four corners are no longer equilibria, and the heteroclinic cycle $\Lambda$ is ``broken".
Now we are going to show that the system (\ref{eqn:closeloop}) undergoes a heteroclinic bifurcation at $\mu=0$ when certain condition holds, and a limit cycle with the same stability of the corresponding heteroclinic cycle may be generated from $\Lambda$ for $\mu>0$ sufficiently close to $0$.

\begin{defn}\label{delta neighborhood}
The internal $\sigma$-neighborhood of $\Lambda$ is the set of all points in $\mathcal{I}$ that are at a distance less than $\sigma>0$ from $\Lambda$.
\end{defn}

\begin{thm}\label{heteroclinicbifurcation}
Consider system (\ref{eqn:closeloop}).  
 If $a>\theta$,  $b<1$,  $c<\theta$, and $d>1$,   then there exist $\sigma,  \epsilon> 0$, such that for $0<\mu< \epsilon$, there is at most one limit cycle $\Gamma$ in the internal $\sigma$-neighborhood
of $\Lambda$, and $\Gamma$ (if it exists) is stable.
\end{thm}
\begin{pf}
The proof is  straightforward by applying the theorem of the heteroclinic bifurcation \cite[Theorem 2.3.2, 2.3.3]{Luo:1997}. When $a>\theta$,  $b<1$,  $c<\theta$, and $d>1$ hold, which implies $ad>bc$.  Then according to Lemma \ref{heterocliniccycle},
there is a  heteroclinic cycle when $\mu=0$ and it is stable.

In addition, all the traces $T_i$ of the Jacobians of the equilibria  are  negative, while their product $\prod_{i=1}^4 T_i$ is positive. Then according to \cite[Theorem 2.3.3]{Luo:1997}, there exist sufficiently small  $\sigma, \epsilon>0$ such that for $0<\mu<\epsilon$, system (\ref{eqn:closeloop}) has at most one limit cycle in the internal $\sigma$-neighbourhood
of $\Lambda$, and this limit cycle (if it exists) is stable due to the stability of $\Lambda$.
\hfill$\qed$
\end{pf}
When $ad<bc$, according to the second statement of Lemma \ref{heterocliniccycle}, the  heteroclinic cycle is unstable. It is impossible for limit cycles to be generated from this unstable heteroclinic cycle in view of  Lemma \ref{nonclosedorbit}, which claims that there are no closed orbits in $\mathcal{I}$ for any $\mu\geq 0$ when $ad-bc<0$.

Compared with the result of Hopf bifurcation, the conclusion of Heteroclinic bifurcation is relatively weaker  since Theorem \ref{heteroclinicbifurcation} only states that there may exist a unique limit cycle in some neighborhood of $\Lambda$. To study the exact conditions for the existence of the unique limit cycle is  beyond this work.

\begin{figure}[htbp!]
	\centering 
	\subfloat[Large limit cycle]{\includegraphics[width=4.2cm]{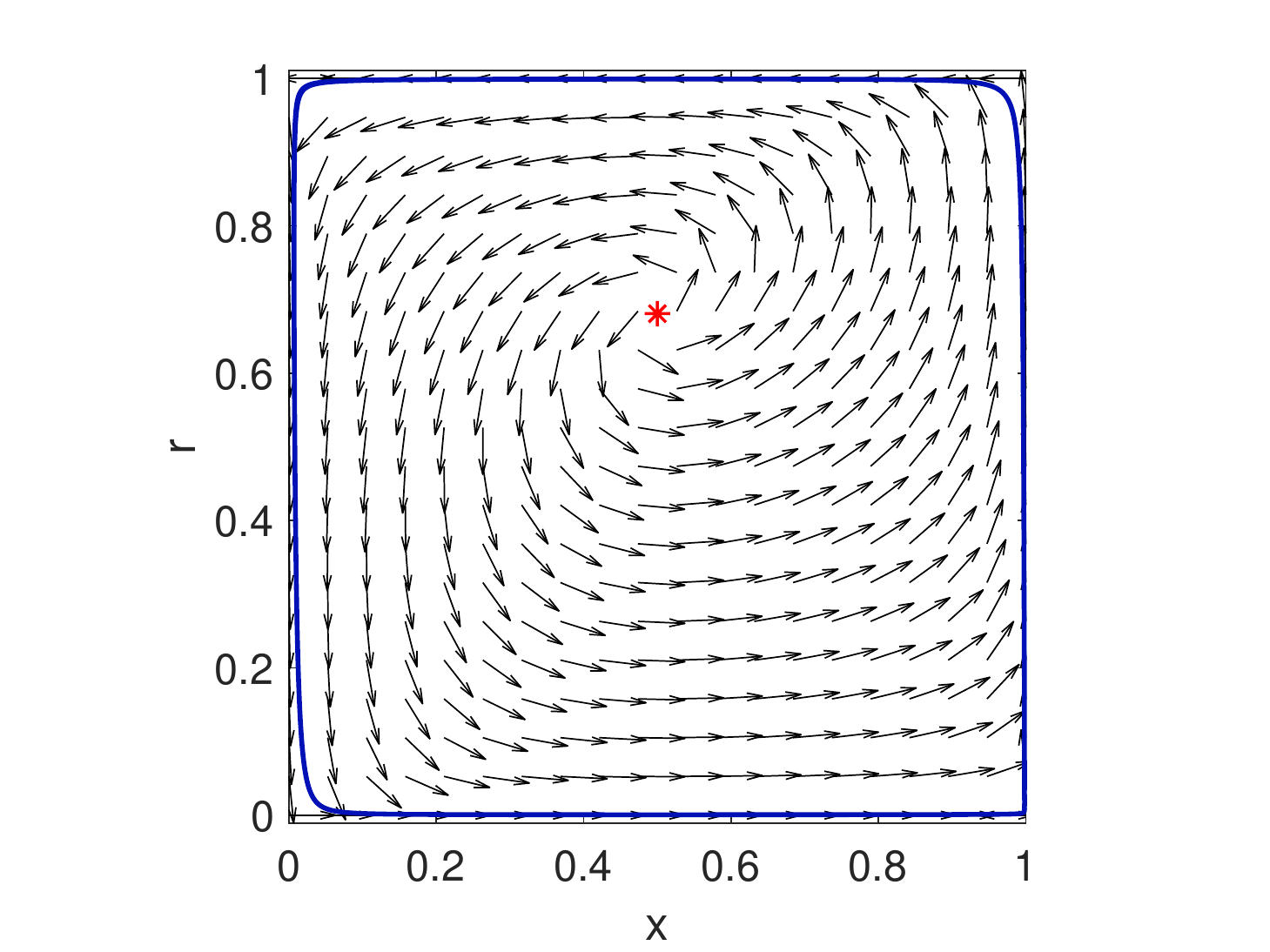}}
	\subfloat[Small limit cycle]{\includegraphics[width=4.2cm]{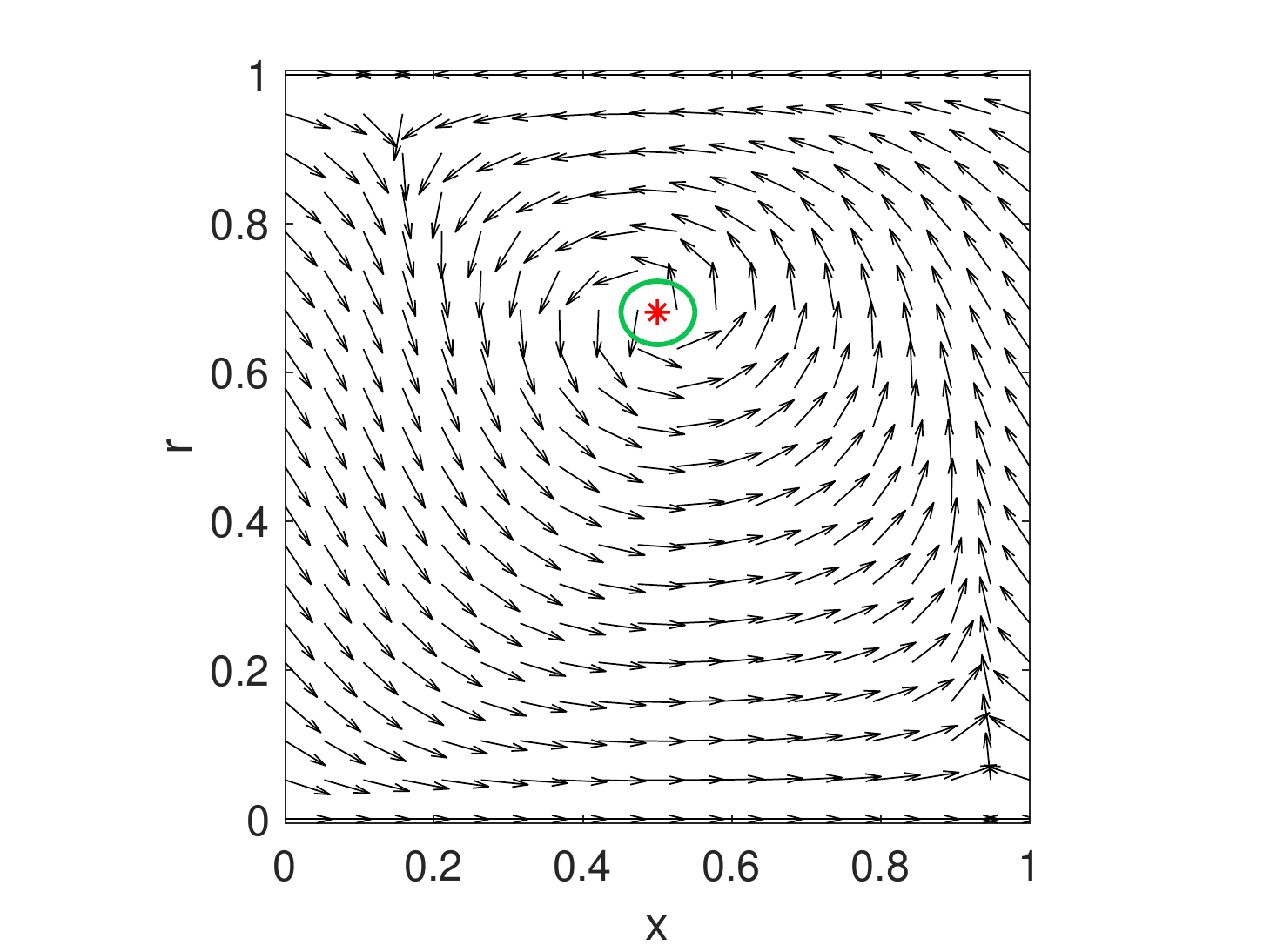}} 
	\caption{Phase portraits of the dynamics in (\ref{eqn:closeloop}) with payoff matrices as shown in (\ref{payoffexample1}). The red stars denote the unique interior equilibrium $(0.5, 0.6809)$. The blue and green cycles show the approximated limit cycles when (a)  $\mu=0.005$; (b)  $\mu=0.1540$.
	} 
	\label{fig:differentlimitcycle}
\end{figure}

Another difference between the limit cycles generated from Hopf bifurcation and Heteroclinic bifurcation is that the limit cycle arising from Hopf bifurcation  is small in size in the phase space, as demonstrated in the simulation results in Fig. \ref{fig:differentlimitcycle}, while the other can be large. In these simulations the payoff matrices are given by
\begin{equation}\label{payoffexample1}
\begin{bmatrix}
R_1&S_1\\
T_1&P_1
\end{bmatrix}=\begin{bmatrix}
3.5&1\\
0.5&0.8
\end{bmatrix},~~
\begin{bmatrix}
R_2&S_2\\
T_2&P_2
\end{bmatrix}=\begin{bmatrix}
2&0.2\\
2.5&1.2
\end{bmatrix}.  \end{equation}
The parameter $\theta$ is chosen to be $1$, and thus the interior equilibrium is fixed and independent of $\mu$. In view of Theorem  \ref{thm:hopfbifurcation},  the Hopf bifurcation point is $\mu_1=0.1543$. 

\subsection{Global dynamics}

As stated before, the limit cycle  analysis for Hopf and Heteroclinic bifurcations are limited to the vicinity of the bifurcation points since the used methods rely mainly on linearization. To study if the limit cycle persists for $\mu\in (0,\mu_1)$, it is necessary to analyze the dynamics of (\ref{eqn:closeloop})  for all such $\mu$. 

\subsubsection{Boundary equilibria and their stability}

Recall that the condition (\ref{condition1}), which involves $\mu$, guarantees that there is a unique interior equilibrium. For some given payoff parameters and fixed values of $\theta$, there may be some $\mu$ such that there is no interior equilibrium. In this case, the system dynamics are relatively trivial since the trajectories will converge to some limit sets on the boundary according to the Poincar\'e-Bendixson theorem. In the following part, we concentrate on the situation where an interior equilibrium always exists for all $\mu\in [0,1]$.  Then, we have the following assumption.
\begin{assum}\label{assumption}
We assume hereafter that $
-(\theta a+\theta^2 b)<(\theta-1)(\theta+1)^2<\theta c+\theta^2 d $.
\end{assum}
Assumption \ref{assumption} ensures that there is a unique interior equilibrium point for all $\mu\in [0,1]$ in system (\ref{eqn:closeloop}). It is noted that when $\theta=1$, Assumption \ref{assumption} is satisfied trivially.

It is easy to check that the boundary $\partial \mathcal{I}: \mathcal{B}_t \cup \mathcal{B}_b\cup \mathcal{B}_l \cup \mathcal{B}_r$ itself and each of its sides are invariant under (\ref{eqn:closeloop}) in the case $\mu=0$. When $0<\mu\leq 1$, the sides  $\mathcal{B}_t$ and $\mathcal{B}_b$ remain invariant, while  $\mathcal{B}_l$ and $\mathcal{B}_r$  do not.
As $\mu$ varies, all the possible boundary equilibria lie on the sides $\mathcal{B}_t$ and $\mathcal{B}_b$, but the positions of these equilibria change. Let us try to determine the locations of the equilibria on $\mathcal{B}_t$ and $\mathcal{B}_b$.  

\begin{lem} \label{lemma9}
 There is only one equilibrium on each side $\mathcal{B}_b$ and $\mathcal{B}_t$ in  system (\ref{eqn:closeloop}). Moreover, the equilibria are located in the sets $\{(x,r): x\in(\max\{1/2,1/(\theta+1)\},1),~r=0\}$ and $\{(x,r) : x\in(0, \min\{1/2,1/(\theta+1)\}),~r=1\}$ respectively.
\end{lem}
\begin{pf}
See Appendix \ref{boundaryequilibrium}.
\hfill$\qed$
\end{pf}

One can examine the stability of these boundary equilibria by evaluating the corresponding Jacobian matrices. 
\begin{lem}\label{lemma10}
 All the boundary equilibria of system (\ref{eqn:closeloop}) are unstable for $\mu \in(0,1]$.
\end{lem}
\begin{pf}
We only show the case on the side  $\mathcal{B}_b$. For the case of $\mathcal{B}_t$, one can obtain the same result analogously.
Denote the equilibrium  on the side $\mathcal{B}_b$ by $(x_b^*,0)$. We obtain its Jacobian matrix as below
\begin{equation}
    J_b^*=\begin{bmatrix}
    \begin{aligned}
    &3x_b^{*^2}(b-a)-2\mu+b\\
    &+2x_b^*(a-2b)
    \end{aligned}
    &\star\\
    0&(\theta+1)x_b^*-1
    \end{bmatrix},
\end{equation}
where the symbol $\star$ stands for $-x_b^{*^3}(-c+d-a+b)-x_b^*(d+b)+x_b^{*^2}(-c+2d-a+2b)$. As the matrix is upper triangular, the entry $(\theta+1)x_b^*-1$ is one of its eigenvalues.
According to Lemma \ref{lemma9}, one has $x_b^*\in (\max\{1/2,1/(\theta+1)\},1)$. Then the eigenvalue $(\theta+1)x_b^*-1$ will be positive for any  $\theta>0$, which implies that the equilibrium is unstable. \hfill$\qed$
\end{pf}

By checking the sign of $\dot{x}$ on the left side $\mathcal{B}_l$ and the right side $\mathcal{B}_r$, one observes that it is always positive and negative respectively. It can be easily verified that the vectors on these two sides point inwards to $\mathrm{int} ( \mathcal{I})$.
The sides $\mathcal{B}_b$ and $\mathcal{B}_t$ are positively invariant under system (\ref{eqn:closeloop}), and thus
the dynamics on each side of $\mathcal{B}_b$ and $\mathcal{B}_t$ are easy to analyze.
For example, on  the side $\mathcal{B}_b$, one has $\dot{r}=0$, $\dot{x}>0$ at $x=0$ and $\dot{r}=0$, $\dot{x}<0$ at $x=1$.  As   there is only one equilibrium $(x_b^*,0)$ on the side $\mathcal{B}_b$, one can easily verify that $\dot{x}>0$ for $x \in [0,x_b^*)$ and $\dot{x}<0$ for $x \in (x_b^*,1]$. Thus, trajectories starting from $\mathcal{B}_b \setminus (x_b^*,0)$ will converge to the unique equilibrium $(x_b^*,0)$. The similar result can be obtained for the dynamics on $\mathcal{B}_t$.

In addition, we can show that the boundary $\partial \mathcal{I}$ is repelling for $\mu \in (0,1]$.

\begin{lem}\label{lem:repelling}
  The boundary $\partial \mathcal{I}$ of system (\ref{eqn:closeloop}) is repelling for $\mu \in (0,1]$. 
\end{lem}
\begin{pf} See Appendix \ref{proof:repelling}.
\hfill$\qed$

\end{pf}

With Lemma \ref{lem:repelling} in hand, we are ready  to present some global results of the system dynamics in $\mathcal{I}$.
\begin{thm}\label{globalstableinteriorequilibrium}
Under Assumption \ref{assumptionofallparameters}, for $\mu \in (\mu_0, 1]$, the interior equilibrium $(x^*,r^*)$ of system (\ref{eqn:closeloop})  is asymptotically stable, and all trajectories starting from  $ \mathcal{I}\setminus (\mathcal{B}_t\cup \mathcal{B}_b)$ will converge to $(x^*,r^*)$.
\end{thm}
\begin{pf}
Recall the two critical parameter values $\mu_0$ and $ \mu_1$, which appear respectively in Lemma \ref{nonclosedorbit} and Theorem \ref{thm:hopfbifurcation}.
One can check that $\mu_0\geq \mu_1$ when Assumption \ref{assumptionofallparameters} holds.   Again according to Lemma \ref{nonclosedorbit}, it is impossible for system (\ref{eqn:closeloop}) to have  closed orbits in $\mathcal{I}$ when $\mu>\mu_0$. In addition, the interior equilibrium is locally asymptotically stable in view of (\ref{interiorJacobian1}). Due to the fact that the boundary $\partial \mathcal{I}$ is repelling as shown in Lemma \ref{lem:repelling},  any trajectories starting from $\text{int}(\mathcal{I})\setminus \{(x^*,r^*)\}$ cannot converge to $\partial \mathcal{I}$. Then according to the Poincar\'e-Bendixson theorem, because of the non-existence of closed orbits or other equilibria in $\text{int}(\mathcal{I})$, the trajectories with any initial points in $ \text{int}(\mathcal{I})\setminus \{(x^*,r^*)\}$ will converge to $(x^*,r^*)$. Similarly, trajectories starting from the sides $\mathcal{B}_l$ and $\mathcal{B}_r$ also converge to $(x^*,r^*)$.
\hfill$\qed$
\end{pf}
So the interior equilibrium is almost globally stable in system (\ref{eqn:closeloop}), because almost all trajectories in $\mathcal{I}$ (except for those starting from $\mathcal{B}_t\cup \mathcal{B}_b$) converge to the stable equilibrium $(x^*,r^*)$. We apply similar terminology   to the stability of the limit cycle when it is stable and almost all trajectories converge to it.

\subsubsection{Uniqueness of limit cycles in the case $\theta=1$}
 \begin{assum}\label{assumption2}
We assume that $\theta=1$ henceforth. 
\end{assum}
Under Assumption \ref{assumption2}, the enhancement effect and the degradation effect are balanced. In this case, the system can be transformed to a generalized Li\'enard system. However, the existence and  uniqueness of limit cycles for $\mu \in (0, \mu_1]$ in the case $\theta\neq 1$  is still open.
\begin{lem}\label{lem:limitcycle}
Under Assumption \ref{assumptionofallparameters},
when $a+c=b+d$, the system dynamics in (\ref{eqn:closeloop}) admit at most one limit cycle in $\text{int}(\mathcal{I})$ for $\mu\in (0,\mu_1)$.
\end{lem}
\begin{pf}
Under Assumption \ref{assumption2}, when $a+c=b+d$, the interior equilibrium,  denoted by $q^*=(1/2,(a+b)/(2(b+d)))$, is fixed.
And the system dynamics (\ref{eqn:closeloop}) are reduced to be
\begin{equation}\label{eqn:closeloop3}
\begin{cases}
&\begin{aligned}
   \dot{x}=&x(1-x)[x(d-c)-r(d+b)+b]+\mu(1-2x)
\end{aligned}\\
&\dot{r}=r(1-r)(2x-1).
\end{cases}
\end{equation}
We apply the transformation of the coordinates on (\ref{eqn:closeloop3}), $x\rightarrow x', ~r\rightarrow 1-r'$, and obtain the following system
\begin{equation*}\label{eqn:closeloop4}
\begin{cases}
&\begin{aligned}
   \dot{x}'=&x'(1-x')[x'(d-c)+r'(d+b)-d]\\
   &+\mu(1-2x')
\end{aligned}\\
&\dot{r}'=-r'(1-r')(2x'-1).
\end{cases}
\end{equation*}
The interior equilibrium $q^*$ of  (\ref{eqn:closeloop3}) now becomes $q'^*=(1/2,r'^*)$ with $r'^*=(a+b)/(2(b+d))$. By using the change of variables $x'\rightarrow \tilde{x}+1/2$ and $r'\rightarrow \tilde{r}+r'^*$, we can shift the equilibrium $q'^*$ to the origin of a new system as below
\begin{equation}\label{eqn:closeloop5}
\begin{cases}
&\begin{aligned}
   \dot{\tilde{x}}=&\alpha(\tilde{x}) \left[(d+b)\tilde{r}+\frac{8\mu_1\tilde{x}\alpha(\tilde{x})-2\mu \tilde{x}}{\alpha(\tilde{x})} \right]
\end{aligned}\\
&\dot{\tilde{r}}=-2\beta(\tilde{r})\tilde{x},
\end{cases}
\end{equation}
which is defined in the region $\tilde{\mathcal{I}}=\{(\tilde{x},\tilde{r}):\tilde{x}\in(-1/2,1/2), \tilde{r}\in (-\tilde{r}^*,1-\tilde{r}^*) \}$ with $\alpha(\tilde{x})=(1/4-\tilde{x}^2)$ and $\beta(\tilde{r})=(\tilde{r}+r'^*)(1-\tilde{r}-r'^*)$. The system is now in the form of the \emph{generalized Li\'enard system}  \cite{Sabatini:06}.

 One can  observe that the functions $\alpha(\tilde{x})$ and $\beta(\tilde{r})$  are always positive in the domain.  Multiplying the vector field of (\ref{eqn:closeloop5}) by the positive function $1/(\alpha(\tilde{x})\beta(\tilde{r}))$, we obtain 
 \begin{equation}\label{eqn:closeloop6}
\begin{cases}
     &\dot{\tilde{x}}=\varphi(\tilde{r})-F(\tilde{x},\tilde{r})\\
     &\dot{\tilde{r}}=-g(\tilde{x}),
\end{cases}
 \end{equation}
where $\varphi(\tilde{r})=\frac{(b+d)\tilde{r}}{\beta(\tilde{r})}$, $F(\tilde{x},\tilde{r})=\frac{2\mu \tilde{x}-8\mu_1\tilde{x}\alpha(\tilde{x})}{\alpha(\tilde{x})\beta(\tilde{r})}$,
 and $g(\tilde{x})=\frac{2\tilde{x}}{\alpha(\tilde{x})}$.  The point-wise direction of the vector field   of (\ref{eqn:closeloop5}) is maintained in (\ref{eqn:closeloop6}).
Hence the existence and uniqueness of limit cycles for the two systems (\ref{eqn:closeloop5}) and (\ref{eqn:closeloop6}) are equivalent.

Under Assumption \ref{assumptionofallparameters}, if $0<\mu<\mu_1$ holds, it can be validated that there always exists a $\nu=\sqrt{\frac{\mu_1-\mu}{4\mu_1}}$ such that the following conditions are satisfied:
\begin{enumerate}
\item $G(-\nu)=G(\nu)$ where $G(\tilde{x}):=\int_{0}^{\tilde{x}} g(s)ds$;
\item $\forall \tilde{x}\in(-\nu,0)$, the function $\tilde{r} \mapsto \frac{F(\tilde{x},\tilde{r})}{\varphi(\tilde{r})}$ is strictly decreasing both on $0<\tilde{r}<1-r'^*$ and $-r'^*<\tilde{r}<0$; $\forall  \tilde{x}\in(0,\nu)$, the function $\tilde{r} \mapsto \frac{F(\tilde{x},\tilde{r})}{\varphi(\tilde{r})}$ is strictly increasing both on $0<\tilde{r}<1-r'^*$ and $-r'^*<\tilde{r}<0$;
\item $\forall \tilde{x}\in (-\nu,\nu)$, $\forall \tilde{r} \in (-r'^*,1-r'^*)$, one has $g(\tilde{x})F(\tilde{x},\tilde{r})\leq 0$;
\item $\forall \tilde{x}\notin (-\nu,\nu)$, $\forall \tilde{r} \in (-r'^*,1-r'^*)$, one has $F(\tilde{x},\tilde{r})\geq 0$; $\forall \tilde{r}\in (-r'^*,1-r'^*)$, the function $\tilde{x} \mapsto F(\tilde{x},\tilde{r})$ is increasing both on $-1/2<\tilde{x}<-\nu$ and $\nu<\tilde{x}<1/2$.
\end{enumerate}
 The detailed expressions of the functions involved in the above conditions can be found in Appendix \ref{detailedexpressions}. It is not difficult to check these conditions, so the process is omitted here.

Then according to \cite[Theorem 1, Corrollary 1]{Sabatini:06}, system (\ref{eqn:closeloop6}) has at most one limit cycle in $\tilde{\mathcal{I}}$. Therefore, the original system  (\ref{eqn:closeloop3}) admits at most one limit cycle in $\text{int}(\mathcal{I})$.
 \hfill$\qed$
\end{pf}

It is noted that when $a+c=b+d$, the coefficient of the cross term $xr$ is $0$, and thus this cross term disappears in the system equations of (\ref{eqn:closeloop3}). So the nonlinearity of this system is reduced. In this case, when system (\ref{eqn:closeloop3}) is transformed into the Li\'enard form, the properties $(1)-(4)$ listed in the proof of Lemma \ref{lem:limitcycle} allow us to use the  Li\'enard system theory to prove that there is at most one limit cycle in $\text{int}(\mathcal{I})$ for $\mu\in (0,\mu_1)$. Unfortunately, those properties are not satisfied when $a+c\neq b+d$, and thus the existence and the number of limit cycles in this case remain unclear.

\begin{figure}[htbp!]
    \centering
\includegraphics[width=6.5cm]{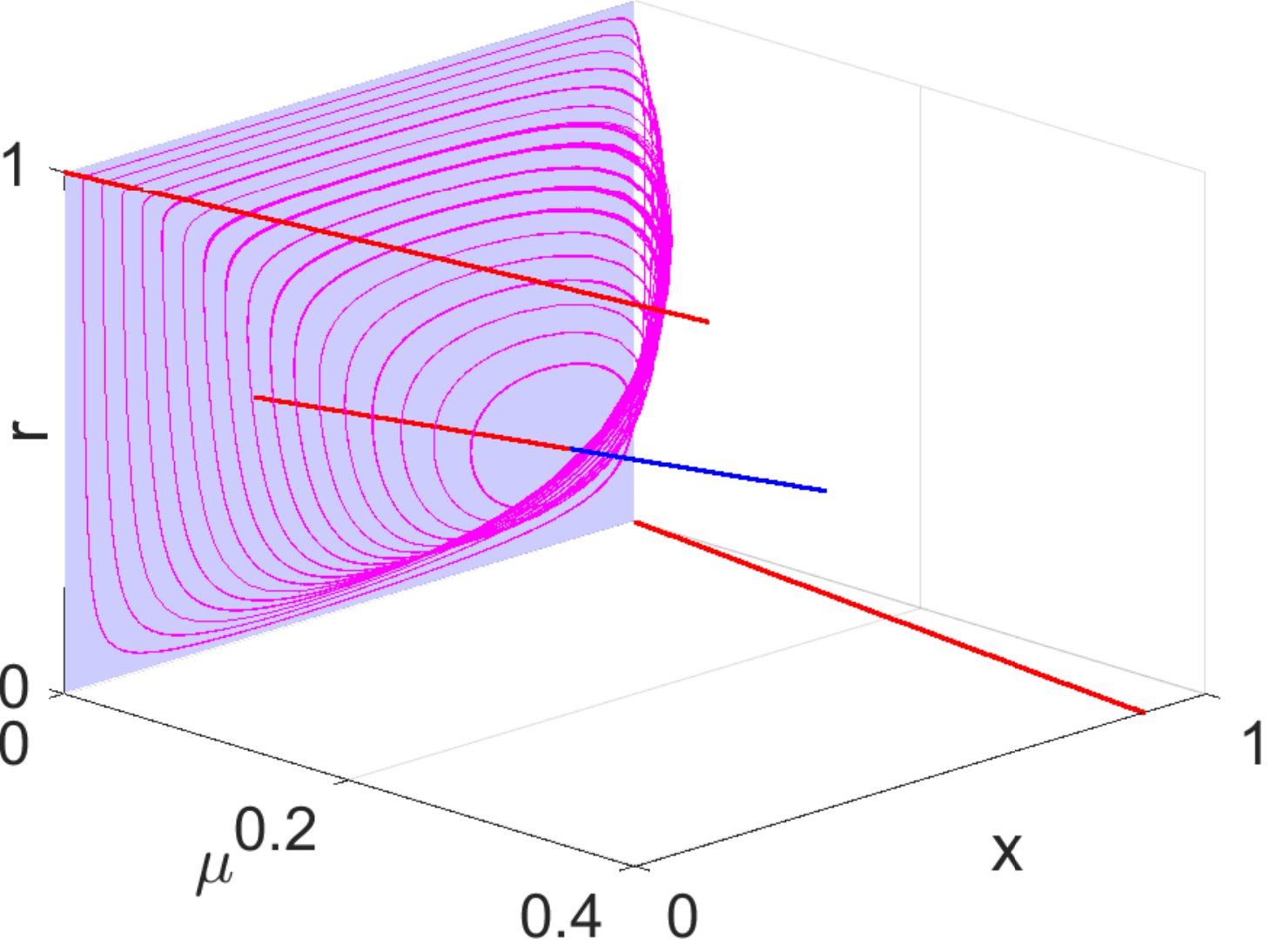}
    \caption{The plot of a unique limit cycle of dynamics (\ref{eqn:closeloop}) with the given payoff matrices in (\ref{payoffmatrixexample2}). The $x$-axis  is the mutation rate $\mu$, the blue and red curves are stable and unstable equilibria respectively, and the magenta curves show the shape of stable limit cycles.
    }
    \label{fig:bifurcationplot}
\end{figure}

Now we are in the position to present the last result of this section about the uniqueness of limit cycle and stability of the interior equilibrium. 
\begin{thm}\label{uniqueness}
 Under Assumption \ref{assumptionofallparameters}, when $a+c=b+d$, 
 \begin{enumerate}
     \item for $\mu\in(0,\mu_1$),  system (\ref{eqn:closeloop}) has exactly one limit cycle that is  almost globally stable;\\
     \item for $\mu\in[\mu_1,1]$, the interior equilibrium  $(x^*,r^*)$ is  almost globally stable.
 \end{enumerate}
 
\end{thm}
\begin{pf}
Under Assumption \ref{assumptionofallparameters}, when $a+c=b+d$, according to Lemma \ref{lem:limitcycle}, the system (\ref{eqn:closeloop}) has at most one limit cycle for $0<\mu<\mu_1$. 
 As shown in Theorem \ref{thm:hopfbifurcation}, the eigenvalues of the Jacobian at $(x^*,r^*)$ have positive real parts, implying that it is repulsive. Then all trajectories starting in the small neighborhood of  this equilibrium (excluding the equilibrium) will diverge, and the existence of infinite many periodic orbits in the neighborhood is excluded.  From Lemma \ref{lem:repelling}, one also knows that the boundary $\partial \mathcal{I}$ is repelling.  According to the Poincar\'e-Bendixson theorem, for any initial states in $\text{int}(\mathcal{I})\setminus \{(x^*,r^*)\} $, its $\omega$-limit set must be a closed orbit. If there are no limit cycles in $\text{int}(\mathcal{I})\setminus \{(x^*,r^*)\}$, then every trajectory starting from $\text{int}(\mathcal{I})\setminus \{(x^*,r^*)\}$ will form a periodic orbit, and an infinite number of periodic orbits will fill $\text{int}(\mathcal{I})\setminus \{(x^*,r^*)\}$.  This contradicts the repulsiveness of the interior equilibrium and  the boundary \cite{Sabatini:06}.

Suppose in $\text{int}(\mathcal{I})\setminus \{(x^*,r^*)\}$, there is some connected region which consists of an infinite number of neutral periodic orbits and is enclosed by two periodic orbits denoted by $\Gamma_1$ and $\Gamma_2$. Then $\Gamma_1$ and $\Gamma_2$ must be limit cycles, which contradicts Lemma \ref{lem:limitcycle} regarding the maximal number of limit cycles.
Therefore, one can conclude that there is exactly one stable limit cycle in $\mathcal{I}$, and it is attractive for $ \mathcal{I}\setminus (\mathcal{B}_t\cup \mathcal{B}_b\cup(x^*,r^*))$.

Note that  when $a+c=b+d$ and $\theta= 1$, we have $\mu_0=\mu_1$. Thus, the second claim is easy to verify according to Theorems \ref{thm:hopfbifurcation} and \ref{globalstableinteriorequilibrium}.
\hfill$\qed$
\end{pf}
Fig. \ref{fig:bifurcationplot} shows the plot of the unique limit cycle of system (\ref{eqn:closeloop}) when $\mu$ changes from $0$ to $0.4$. The  payoff matrices used in the simulations are shown below, and they satisfy the condition $a+c=b+d$.
\begin{equation}\label{payoffmatrixexample2}
\begin{bmatrix}
R_1&S_1\\
T_1&P_1
\end{bmatrix}=\begin{bmatrix}
3.5&2\\
0.5&1
\end{bmatrix},~~
\begin{bmatrix}
R_2&S_2\\
T_2&P_2
\end{bmatrix}=\begin{bmatrix}
2&0.2\\
3&3.2
\end{bmatrix}.  
\end{equation}
The Hopf bifurcation point is computed to be $\mu_1=0.1633$.
It can be observed that the limit cycle ``collides" with the heteroclinic cycle lying on the boundary and the interior equilibrium as $\mu$ approaches $0$ and $0.1633$ respectively. In addition, it disappears after $\mu$ passes $0.1633$.

\section{Incentive-based Control}
In  real biological and social systems, the mutation or exploration rate  is usually small. For example, an extremely high  mutation rate for a biological entity may reach $1/400$ per site as reported in \cite{Gago1308}.
As known from Section 3,
although the Hopf bifurcation point $\mu_1$ depends on the payoff parameters,   limit cycles exist in system (\ref{eqn:closeloop})  for $\mu<\mu_1$. It means that in many situations stable limit cycles persist in the  co-evolutionary dynamics of the game and environment. On the one hand, the shapes and positions of limit cycles in  nonlinear systems are generally difficult to identify analytically; on the other hand, such limit cycle oscillations,  although different than  the \emph{(oscillating) tragedy of the commons}, are not desirable from a control point of view.
In addition, considering the public resource management, the issues of how to best govern common-pool
resources have long been a common concern in the field of economics \cite{ostrom2015}. 
Thus, in this section we intend to study how to design suitable control policies such that the following two objectives are attained: 
\begin{enumerate}[(a).]
    \item  alleviate or even eliminate the oscillation;
    \item increase the environmental resource stock.
\end{enumerate}


One needs to choose what the control input should be. Although multiple possible approaches have been proposed in the evolutionary games field, perhaps the most plausible control input is to offer suitable incentives to be added  to the payoff of the desired strategies \cite{Riehl:18}.
Now we consider that in each game interaction, the cooperator receives  certain constant incentive $u\geq 0$ from an external regulating authority.

Recall that the entries on the first row of $A(r)$ are the payoffs to a C player. The external incentive will be added to the first row of  (\ref{eq:matrix}). Thus,
the control input $u$ is incorporated  into  the previously defined payoff matrix (\ref{eq:matrix}) as follows
\begin{equation}\label{payoffmatrixwithcontrolinput1}
\begin{aligned}
&A_u(r)=(1-r)\begin{bmatrix}
R_1&S_1\\
T_1&P_1
\end{bmatrix}+r\begin{bmatrix}
R_2&S_2\\
T_2&P_2
\end{bmatrix}+\begin{bmatrix}
u&u\\
0&0
\end{bmatrix} \\
&=\begin{bmatrix}
&r(R_2-R_1)+R_1+u&r(S_2-S_1)+S_1+u\\
&r(T_2-T_1)+T_1&r(P_2-P_1)+P_1
\end{bmatrix}.
\end{aligned}
\end{equation}
Substituting (\ref{payoffmatrixwithcontrolinput1}) into equations (\ref{eqn:closeloop}) results in the following system with the control input $u$
\begin{equation}\label{eqn:closeloopwithcontrol}
\begin{cases}
&\begin{aligned}
   \dot{x}=&x(1-x)[xr(-c+d-a+b)+x(a-b)\\
   & -r(d+b)+b+u]+\mu(1-2x)
\end{aligned}\\
&\dot{r}=r(1-r)(2x-1).
\end{cases}
\end{equation}
To emphasize the effect of the control input $u$ on the system behaviors, we have assumed that the enhancement effect and the degradation effect are balanced by fixing the parameter $\theta$ to be $1$ (under Assumption \ref{assumption2}).
In the following part we will study whether such a constant incentive can achieve the desired control objectives.
\subsection{Stable equilibrium}
System (\ref{eqn:closeloopwithcontrol}) has a unique interior equilibrium which is given by $(x_c^*,r_c^*)=(1/2, ( a+b+2u)/(a+c+b+d))$ when $0\leq u<(c+d)/2$. It is noted that the interior equilibrium is shifted up vertically, i.e., the $x$-coordinate of this equilibrium is the same as the equilibrium of the system without control, while the value of the $r$-coordinate  increases. Then we evaluate the Jacobian at $(x_c^*,r_c^*)$, which is given by
\begin{equation}\label{interiorjacobian2}
 J_c^*=\begin{bmatrix}
\frac{ad-bc+(b+d-a-c)u}{2(a+b+c+d)}-2\mu&\frac{-(a+b+c+d)}{8}\\
\frac{2(a+b+2u)(c+d-2u)}{(a+b+c+d)^2}&0
\end{bmatrix}.   
\end{equation}
For this matrix,
it is easy to identify that the real parts of its eigenvalues are of the same sign, which is the same as the first entry.

When $u\geq(c+d)/2$, no equilibria exist in the interior of $\mathcal{I}$, and all the equilibria are located on the sides $\mathcal{B}_t$ and $\mathcal{B}_b$. For system (\ref{eqn:closeloopwithcontrol}), if $u\geq 0$, it is easy to identify that the $x$-coordinate of  the equilibrium on $\mathcal{B}_b$ is in the range $x\in(1/2,1)$ by using a similar approach in Lemma \ref{lemma9}. Then the same argument of Lemma \ref{lemma10} is applicable, so one can ensure that this equilibrium is unstable for $\text{int}(\mathcal{I})$. Similarly, on the side $\mathcal{B}_t$, when $0<u<(c+d)/2$, the equilibria are located in the range $x\in (0,1/2)$, and thus they are unstable. However, when $u>(c+d)/2$, the situation will be different, and  we have the following statement for the equilibria.
\begin{lem}\label{equilibriumbu}
Consider system (\ref{eqn:closeloopwithcontrol}). When $u>(c+d)/2$, on the side $\mathcal{B}_t$, there exists one equilibrium, denoted by $(x_t^*,1)$, satisfying $x_t^*\in(1/2,1)$. And the other possible equilibria are all located in the set $\{(x,r) : x\in(0, 1/2),~r=1\}$.
\end{lem}
\begin{pf}
See Appendix \ref{equilibriumxustar}.
\hfill$\qed$
\end{pf}
It is easy to check the stability of these equilibria by evaluating the Jacobian.
\begin{lem}
When $u>(c+d)/2$,
the boundary equilibrium $(x_t^*,1)$ is locally asymptotically stable under system (\ref{eqn:closeloopwithcontrol}), while the other equilibria   (if they exist) are unstable.
\end{lem}
\begin{pf}
One can check the local stability of $(x_t^*,1)$ by examining the Jacobian which is given by
\begin{equation}\label{bjacobian3}
 J_t^*=\begin{bmatrix}
\begin{aligned}
&3(c-d){x_t^*}^2+2(2d-c-u)x_t^*\\
&-d+u-2\mu
\end{aligned}&\bullet\\
0&1-2x_t^*
\end{bmatrix},  
\end{equation}
where the symbol $\bullet$ stands for $-x_t^{*^3}(-c+d-a+b)-x_t^*(d+b)+x_t^{*^2}(-c+2d-a+2b)$. Note that the eigenvalues of $J_t^*$ are actually the two diagonal entries. According to Lemma \ref{equilibriumbu}, obviously the eigenvalue $\lambda_1=1-2x_t^*$ is negative.

For the other eigenvalue $\lambda_2=3(c-d){x_t^*}^2+2(2d-c-u)x_t^*-d+u-2\mu$, one can prove it is negative in all the three cases regarding the relationship of $c$ and $d$. To illustrate, we only show the case of $c>d$ here, since the other cases are similar. Suppose $c=d+\epsilon$ with $\epsilon>0$.  From Appendix \ref{equilibriumxustar}, we have  $1/2<  x_t^*<(u-d)/\epsilon$, then one can check that
\begin{equation*}
\begin{aligned}
\lambda_2&=(3{x_t^*}^2-2x_t^*)\epsilon+(u-d)(1-2x_t^*)-2\mu\\
&< (3{x_t^*}^2-2x_t^*)\epsilon +\epsilon x_t^*(1-2x_t^*)-2\mu\\
&=({x_t^*}^2-x_t^*)\epsilon-2\mu\\
&<-2\mu.
\end{aligned}
\end{equation*}
Thus, the eigenvalue $\lambda_2$ is negative. 
The fact of two negative eigenvalues ensures that the equilibrium $(x_t^*,1)$ is locally asymptotically stable. The instability of other possible equilibria is easy to check since there is always a positive eigenvalue in view of (\ref{bjacobian3}).
\hfill$\qed$
\end{pf}

Then, we define a Dulac function which is similar to that  in Section 3.1 
\begin{equation}\label{dulacfunction2}
\varphi_c(x,r)= x^{\alpha} (1-x)^{\beta} r^{\gamma-0.5u} (1-r)^{\delta+0.5u},
\end{equation}
where the parameters $\alpha$, $\beta$, $\gamma$, and $\delta$ are the same as in (\ref{alphavalue}) with $\theta=1$. Denote the vector field of (\ref{eqn:closeloopwithcontrol}) by $f_c(x,r)$.   After the multiplication of $\varphi_c(x,r)$, we compute the divergence of the modified vector field $\varphi_c  f_c(x,r)$ in  $\text{int}(\mathcal{I})$,  which yields
\begin{equation}\label{divergence3}
\begin{aligned}
 \text{div}~\varphi_c  f_c(x,r)
 &=\varphi_c(x,r)[g(x)\mu-2\mu+(1.5+\alpha)u \\
 &~~~ -\frac{(bc-ad)}{2(a+b+c+d)}],
 \end{aligned}
\end{equation}
 with $g(x)$ being the same as in (\ref{divergence1}), which takes the maximum value at $\bar{x}$ as shown in (\ref{xbar}).
 
 Next we discuss  in two cases and show that there always exists some  $u$ such that system  (\ref{eqn:closeloopwithcontrol}) has a stable equilibrium for the whole interior.

\begin{thm}\label{globalstable}
For system (\ref{eqn:closeloopwithcontrol}), when $a+c> b+d$, there exists some $u_1:=\frac{ad-bc-(4-g(\bar{x}))\mu(a+b+c+d)}{a+c-b-d}$ such that
\begin{enumerate}
    \item  if $u_1<(c+d)/2$,  the interior equilibrium $(x_c^*,r_c^*)$ is almost globally stable  under  $u\in (u_1, (c+d)/2)$;
    \item if $u_1>(c+d)/2$,  the boundary equilibrium $(x_t^*,1)$ is almost globally stable under $u>(c+d)/2$.
\end{enumerate}
\end{thm}
\begin{pf}
When $a+c>b+d$, we have $\alpha=-(1+\frac{a+c}{a+b+c+d})\in(-2,-1.5)$, so the coefficient of the term with $u$ in (\ref{divergence3}), i.e., $1.5+\alpha$, is negative.
Thus,  if
\begin{equation}\label{u1}
u>u_1:=\frac{ad-bc-(4-g(\bar{x}))\mu(a+b+c+d)}{a+c-b-d},   
\end{equation}
the right hand side of (\ref{divergence3}) is always negative, which implies that there are no closed orbits in $\mathcal{I}$.

By checking the Jacobian (\ref{interiorjacobian2}), one can see that the interior equilibrium $(x_c^*,r_c^*)$ is stable when \[u>u_2:=\frac{ad-bc-4\mu(a+b+c+d)}{a+c-b-d}.\] As $g(\bar{x})\geq 0$, we have $u_1\geq u_2$.
Then if $u_1<(c+d)/2$, one can choose some $u\in(u_1,(c+d)/2)$ such that the interior equilibrium $(x_c^*,r_c^*)$ exists and is asymptotically stable.
Due to the non-existence of closed orbits in $\mathcal{I}$ and other stable equilibria, and also because of the repulsiveness of $\partial \mathcal{I}$, one can conclude that almost all trajectories  will converge to $(x_c^*,r_c^*)$ asymptotically. 

On the other hand, if $u_1\geq  (c+d)/2$, one can  choose some $u>(c+d)/2$ which leads to the occurrence of the equilibrium $(x_t^*,1)$. In view of (\ref{bjacobian3}), the equilibrium $(x_t^*,1)$ is locally asymptotically stable. The other equilibria on $\mathcal{B}_t$ and $\mathcal{B}_b$ are all unstable, and they are located on the sets whose sufficient close neighborhoods in $\text{int}( \mathcal{I})$ are repelling. Hence, it is impossible for trajectories starting from $\text{int}( \mathcal{I})$ to converge to these points. As a consequence, $(x_t^*,1)$ is the only $\omega$-limit  set for all initial states in $\text{int}( \mathcal{I})$ from the Poincar\'e-Bendixson theorem. Similarly, this result extends to the sides $\mathcal{B}_l$ and $\mathcal{B}_r$ by following the preceding analysis.
\hfill$\qed$
\end{pf}

\begin{thm}\label{globalstable2}
For system (\ref{eqn:closeloopwithcontrol}), when $a+c\leq b+d$,  the boundary equilibrium $(x_t^*,1)$ is almost globally stable under $u>(c+d)/2$.
\end{thm}
\begin{pf}
When $a+c\leq b+d$,  the first entry of (\ref{interiorjacobian2}) will always be positive for any $u$ if $\frac{ad-bc}{2(a+b+c+d)}-2\mu>0$, in which case the equilibrium will  always be unstable (saddle). Thus,
it is impossible to stabilize the interior equilibrium with any positive $u$ if it is unstable in the uncontrolled system. However, one can choose some $u>(c+d)/2$ such that there is an equilibrium $(x_t^*,1)$ on $\mathcal{B}_t$ which can be proved to be almost globally stable  by using the same argument in the second case of Theorem \ref{globalstable}.
\hfill$\qed$
\end{pf}

In Theorems \ref{globalstable} and \ref{globalstable2}, the relationship between $a+c$ and $b+d$  determines whether the interior equilibrium can be stabilized or not under the positive control input $u$. In view of (\ref{interiorjacobian2}), one can observe that the stability of the interior equilibrium can be changed from unstable to stable when $a+c>b+d$ under a suitable $u$, while the stability is not qualitatively affected by $u$ when $a+c\leq b+d$.

\begin{figure}[htbp!]
	\centering 
	\subfloat[stable int. eq.]{\includegraphics[width=4.2cm]{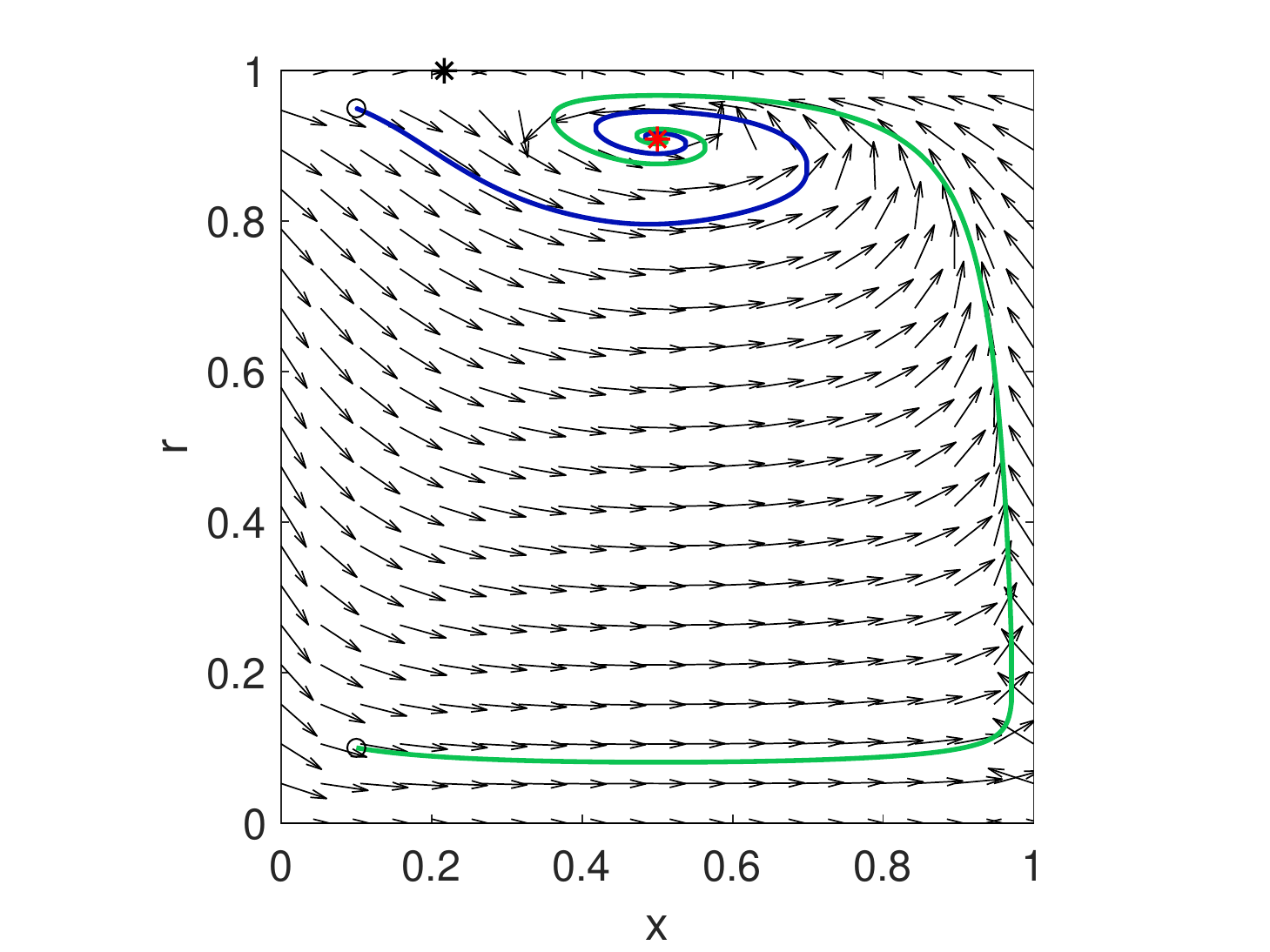}}~~
	\subfloat[stable eq. on $\mathcal{B}_t$]{\includegraphics[width=4.2cm]{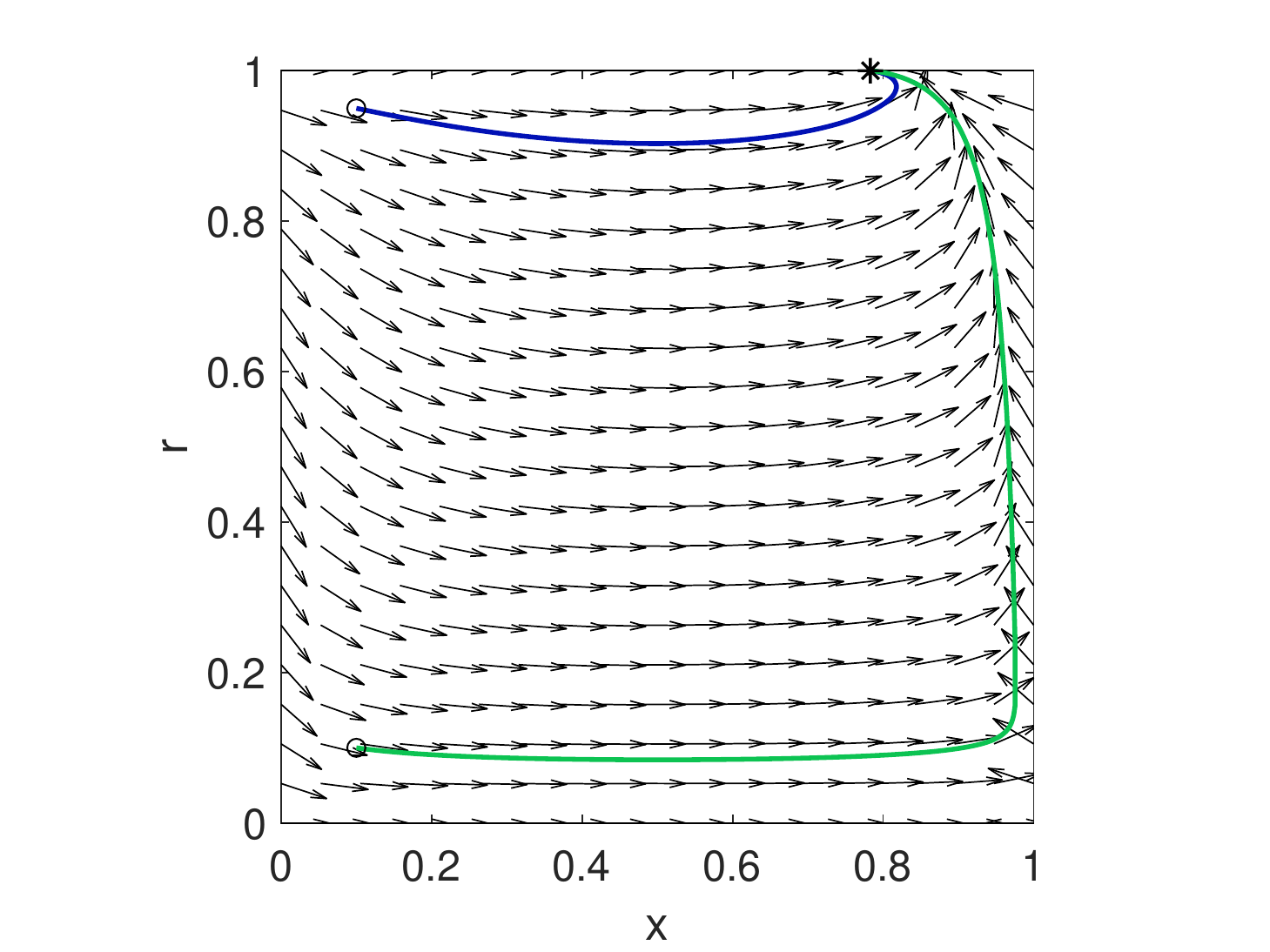}} \\
	\subfloat[unstable int. eq.]{\includegraphics[width=4.2cm]{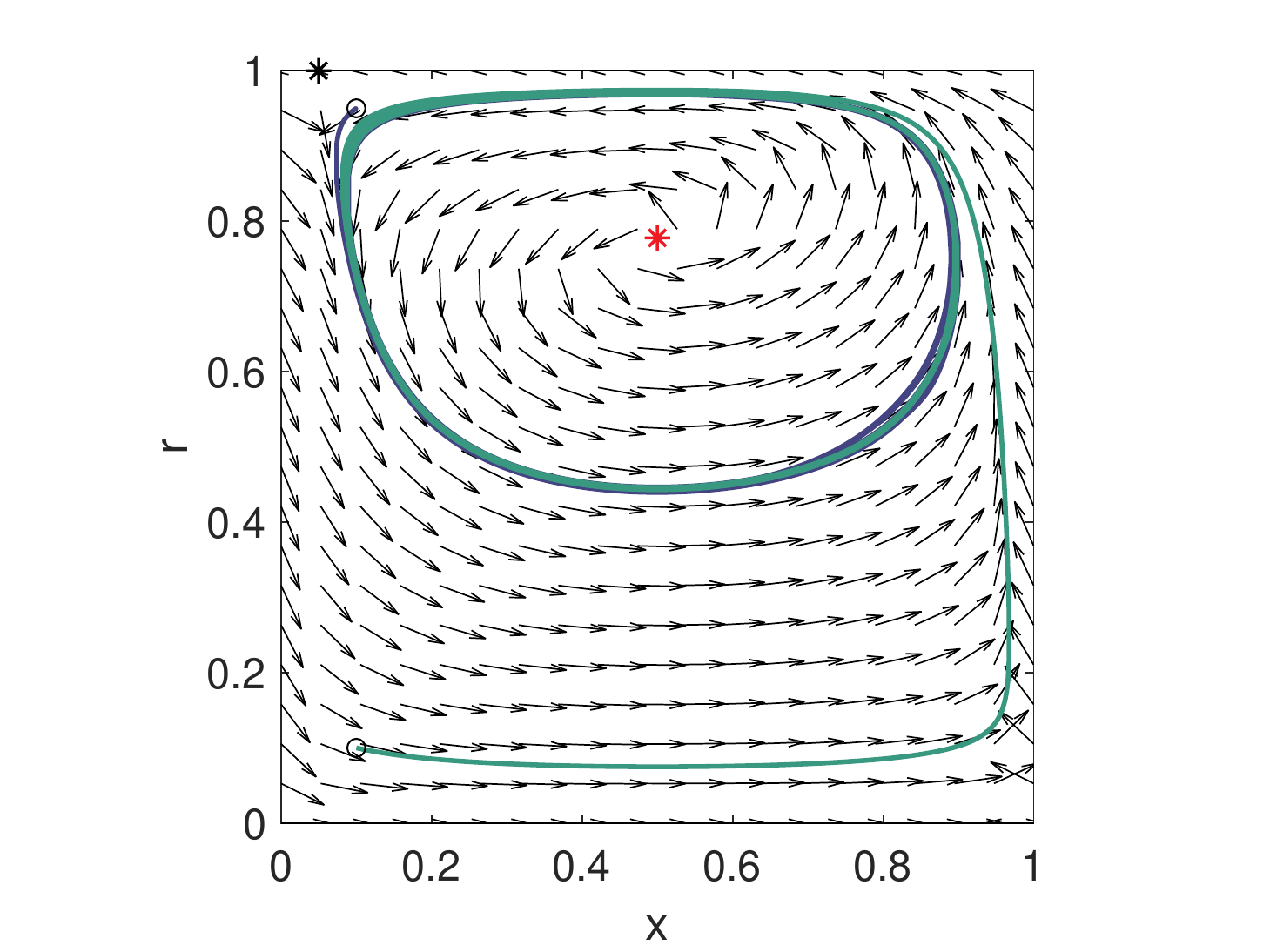}} ~~\subfloat[stable eq. on $\mathcal{B}_t$]{\includegraphics[width=4.2cm]{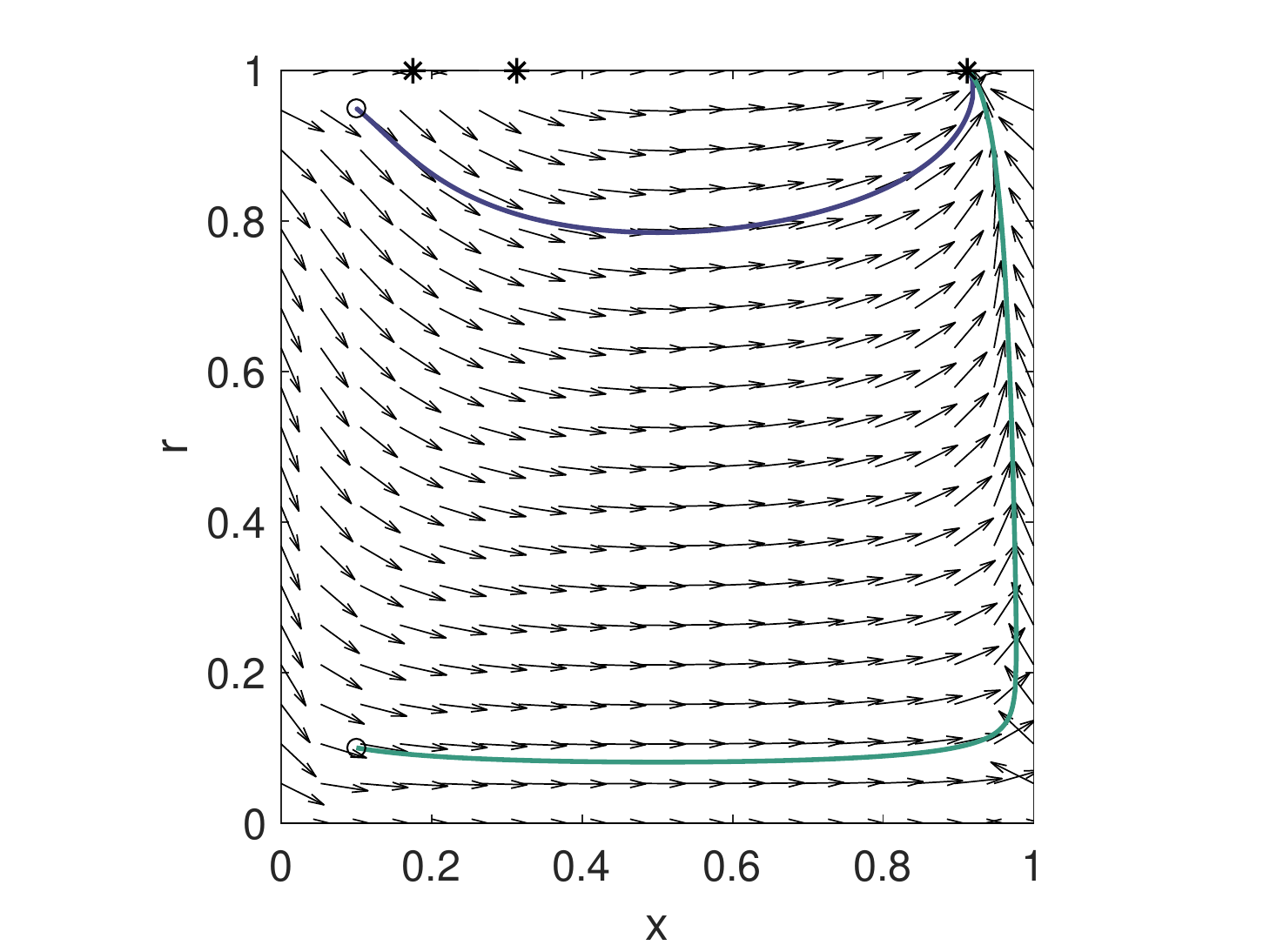}} 
	\caption{Phase portraits of dynamics (\ref{eqn:closeloopwithcontrol}) when $a+c\neq b+d$.  (a) and (b) respectively show the trajectories converge to a stable equilibrium in the interior and on the boundary with different control inputs when $a+c>b+d$ with $a=4$, $b=1$, $c=3$, $d=3$, and $\mu=0.05$. The control inputs in $(a)$ and $(b)$ are $2.8$ and $3.5$ respectively. (c) and (d) are for the case $a+c<b+d$ where $a=2$, $b=3$, $c=1$, $d=4$, and $\mu=0.15$.  In (c) the interior equilibrium cannot be stabilized with small control input with $u=1.8$ , but in (d) one equilibrium on the boundary is stable under a large control input with $u=2.6$.
	} \label{fig:withcontrol}
\end{figure}

When there exists an equilibrium that is globally stable for $\mathcal{I}$,  obviously the limit cycle oscillation is eliminated. No matter whether this stable equilibrium  is in the interior or on the top side. The second designed goal is also achieved, since the $r$-coordinates of the equilibria are higher.
One can find some illustrations in Fig. \ref{fig:withcontrol}. 

\subsection{Reduced amplitude of the limit cycle}
For the case of $a+c=b+d$, although it is impossible to stabilize the interior equilibrium with a suitable $u$, one can observe that  the amplitude of the oscillation on $r$-coordinate  decreases under some small $u$. 

\begin{figure}[htbp!]
	\centering 
	\subfloat[no control input]{\includegraphics[width=4.2cm]{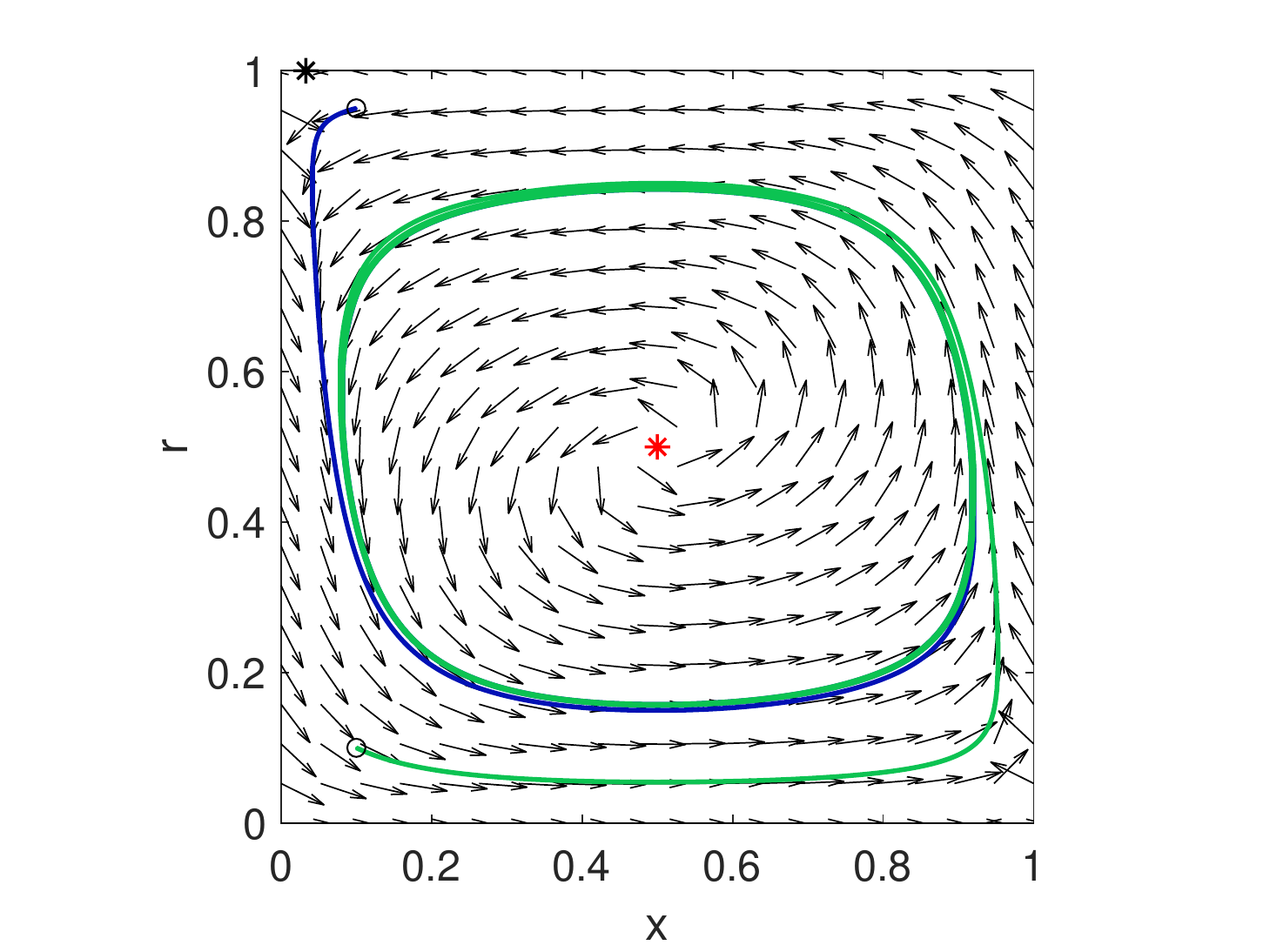}}~~
	\subfloat[intermediate control]{\includegraphics[width=4.2cm]{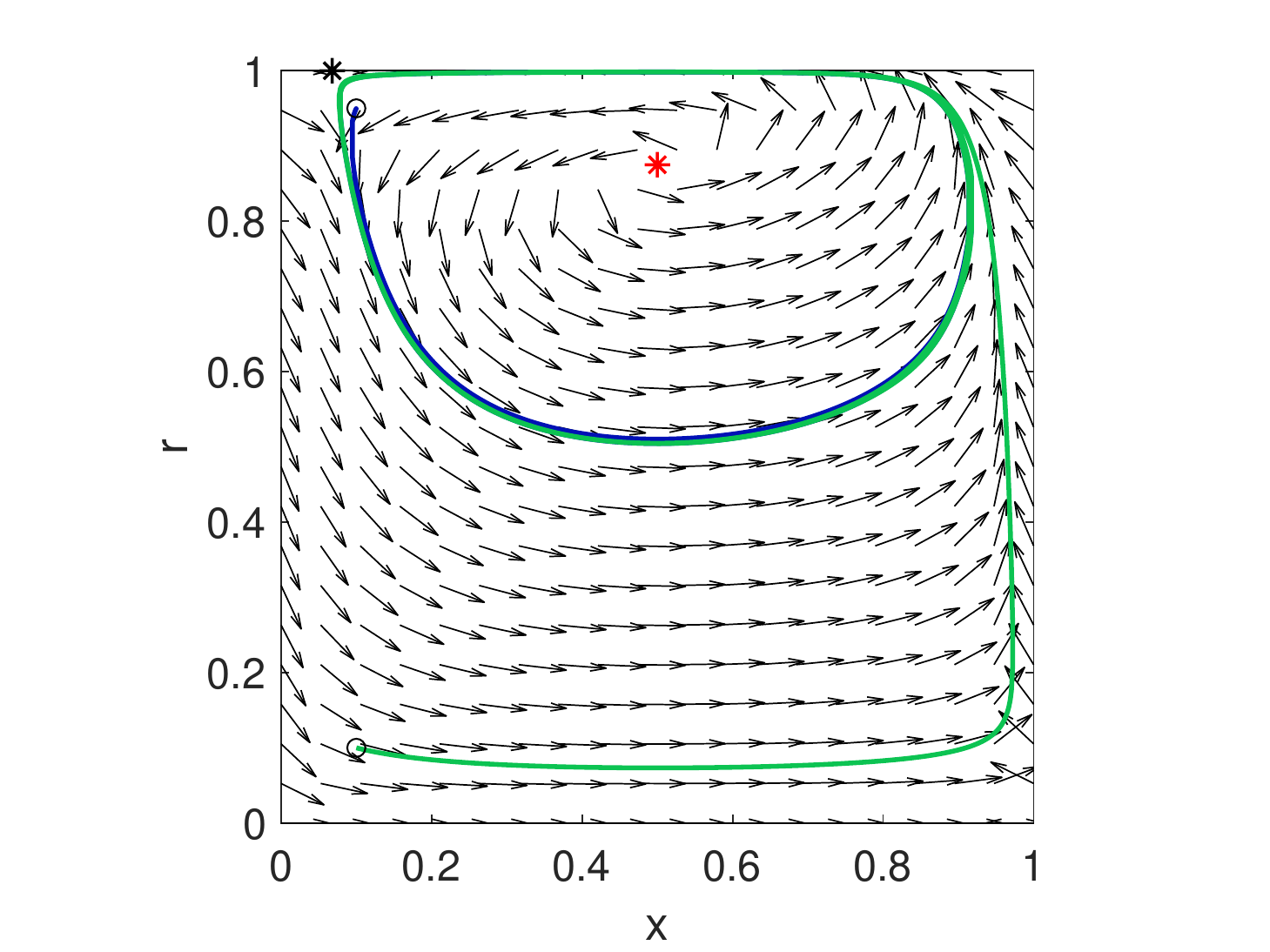}} \\
	\subfloat[stable eq. on $\mathcal{B}_t$]{\includegraphics[width=4.2cm]{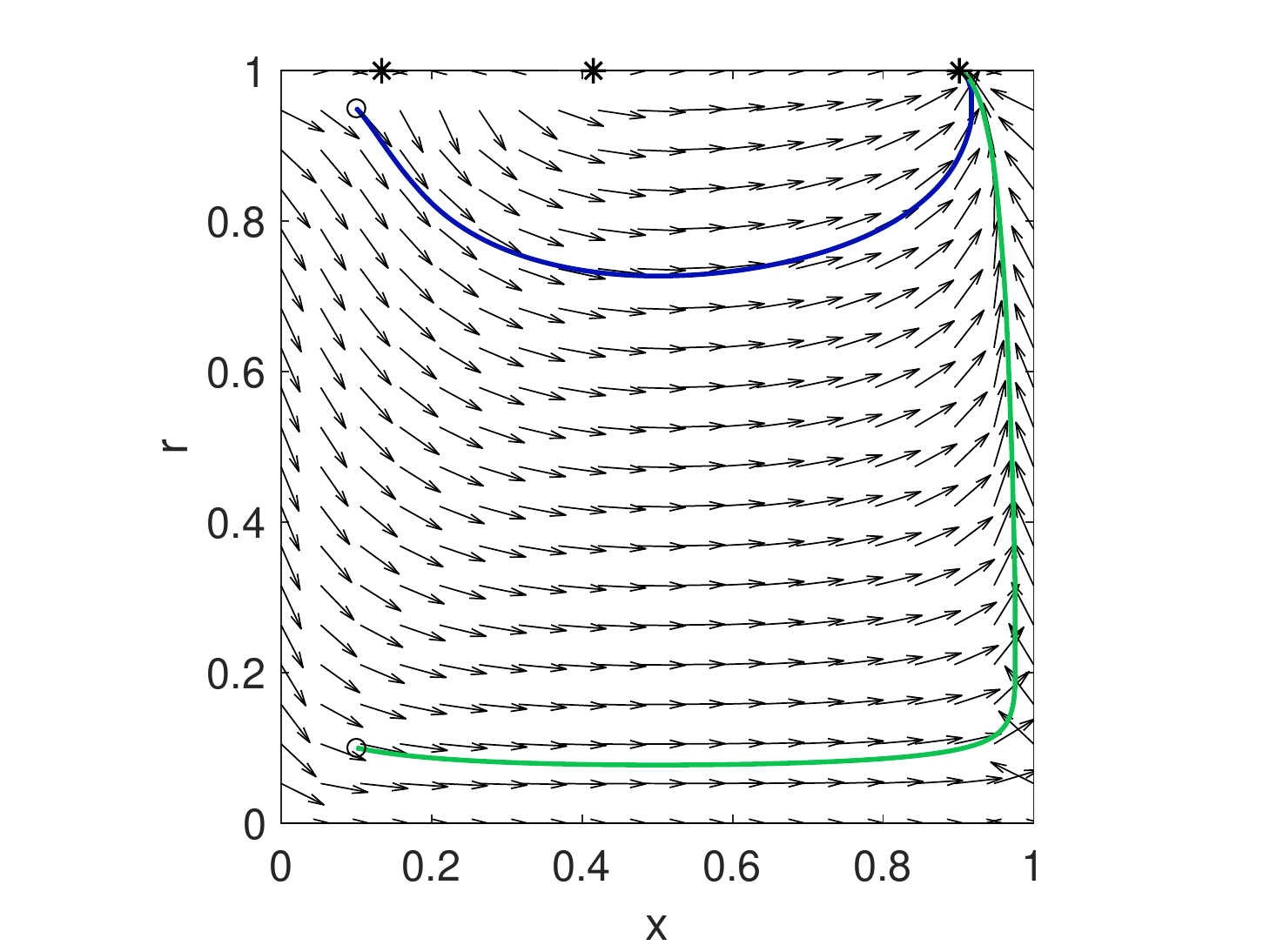}} ~~\subfloat[decreasing amplitude]{\includegraphics[width=4.2cm]{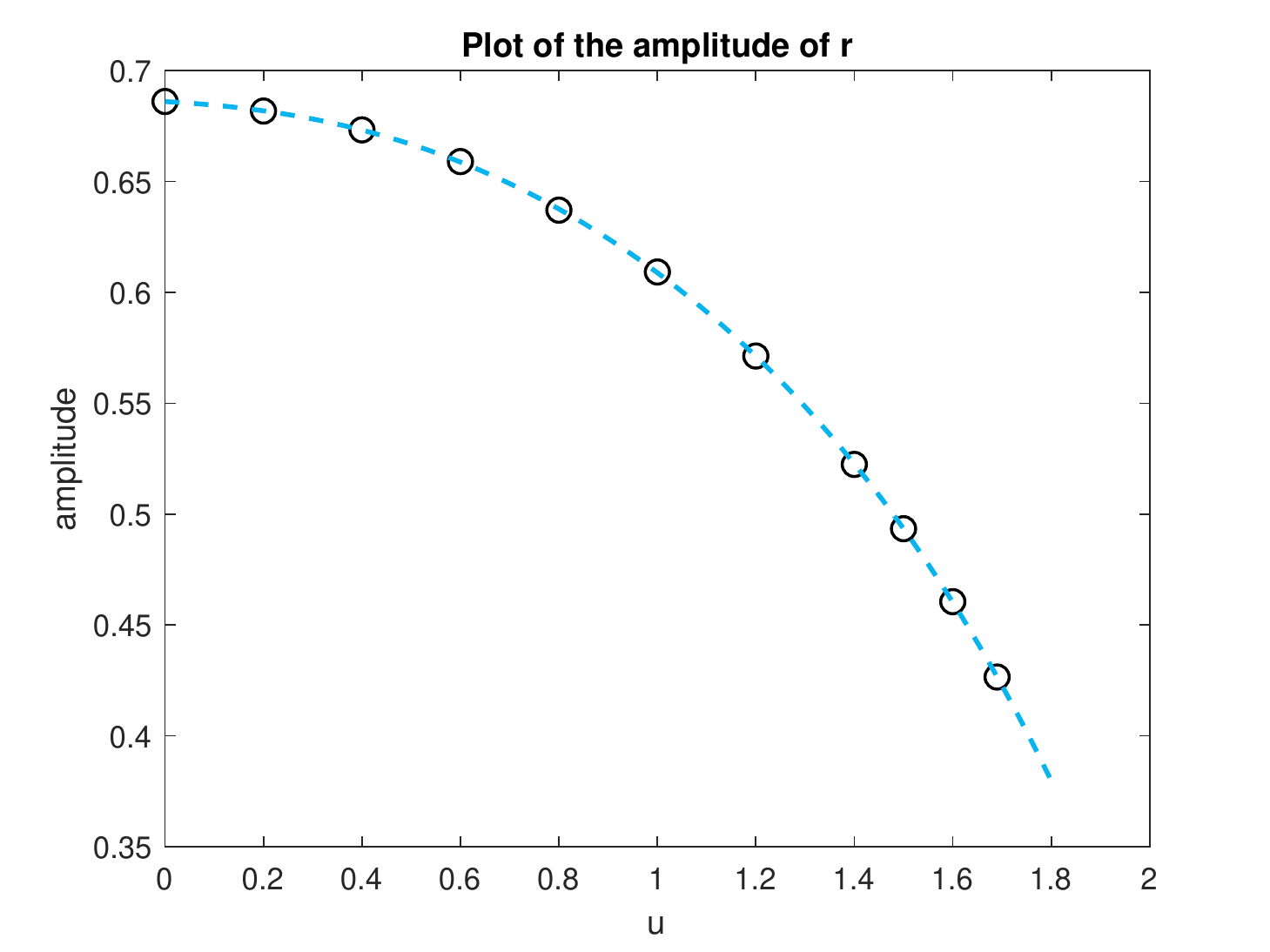}} 
	\caption{Reduced amplitudes of oscillation of variable $r$ when $a+c=b+d$ with $a=3$, $b=1$, $c=1$, $d=3$, and $\mu=0.05$. The first three plots show the phase portraits with different values of $u$: (a) $u=0$; (b) $u=1.6$; (c) $u=2.1$. (d) shows that the amplitude of $r$ of the limit cycle (approximated by the numerical integration)  decreases as $u$ increases.
	} 
	\label{fig:withcontrol1}
\end{figure}

As $a+c= b+d$, it is noted that the control input does not affect the stability of equilibrium $(x_c^*,r_c^*)$ according to the Jacobian (\ref{interiorjacobian2}). However, one can transform system (\ref{eqn:closeloopwithcontrol}) into a form of Li\'enard system  as  (\ref{eqn:closeloop6}) in the same manner. Since the listed conditions are satisfied straightforwardly, according to Lemma (\ref{lem:limitcycle}), 
for $\mu<(0,\mu_1)$, system (\ref{eqn:closeloopwithcontrol}) has a unique limit cycle when $ u\in[0,(c+d)/2)$. When there is some intermediate control input $u\in (0,(c+d)/2)$, the corresponding interior equilibrium is shifted towards the top side $\mathcal{B}_t$. To some extent, the vector field between the interior equilibrium and the top side is ``compressed". 
The position of the limit cycle is shifted up,  and the amplitude  of the cyclic oscillation on $r$-coordinate is reduced at the same time. In this sense, the oscillation is alleviated. Some numerical results can be found in Fig. \ref{fig:withcontrol1}.

\section{Conclusions and Future Work}
We have investigated the co-evolutionary dynamics of the game and environment when strategies' mutations are taken into account. We used the replicator-mutator model to depict the strategic evolution. By using  bifurcation theory we showed that the system at the interior equilibrium undergoes a Hopf bifurcation which generates  stable limit cycles; on the other hand, we showed there is a heteroclinic bifurcation which may also generate a stable limit cycle. Our analysis highlights the important role that mutations play in the integrated system dynamics.
For the co-evolutionary system, the stable limit cycle corresponds to a sustained oscillation of population's decisions and richness of the environmental resource. Compared with the neutral periodic oscillation or heteroclinic cycle oscillation that have been exhibited in previous studies using the replicator model, such  limit cycle oscillations
can better explain the phenomena that appear in real biological or social systems.
Moreover, we proved that under certain parameter conditions, the system always admits a unique stable limit cycle which is attractive for almost the entire domain. 

Such a robust limit cycle oscillation may provide  implications and theoretical support for relevant studies in biology and sociology. For example, in the microbial context, some cooperating bacteria  can produce costly public good, while the defecting bacteria only profit. Thus, the richness of the public good is enhanced by the cooperating bacteria and degraded by the defectors. In the process of bacterial reproduction, mutations are common. The amount of mutant offsprings can have important impact on the dynamic outcomes of the biological systems. For such microbial systems, \cite{brown2007durability} shows that the dynamics can converge to an interior equilibrium, while \cite{Weitz:16} shows that persistent heteroclinic oscillations can exist. 
However, the  results obtained in this work theoretically reveal that a moderate limit cycle oscillation can exist in addition to the convergence to a steady state or a heteroclinic cycle.
Thus, this work complements the previous studies. 

We also considered the incentive-based control  applied into the co-evolutionary dynamics, and showed the effectiveness of such control on maintaining the environmental resource. Namely, compared with the uncontrolled system, the incentive will increase the level of environmental resource in most cases. In particular, one can apply some intermediate incentive to let the system have an interior equilibrium that is stable for the whole interior of domain. If the interior equilibrium cannot be stabilized, some large incentive could be put into use to make the system dynamics converge to a boundary equilibrium with full
environmental resource. 

The obtained results in the control part of this work can provide valuable insights into many practical issues in nature, such as the common-pool resource management, environmental and sustainable development, and ecological diversity conservation. For example, the global climate change is one of the biggest challenges of our times. The
strategic decisions of individuals, corporations, and governments 
have long-term environmental consequence that will, in turn, alter
the strategic landscape those parties face \cite{Tilman:20}.
To mitigate the global warming, all concerned parties are called upon to reduce greenhouse gas emissions. In such a socio-ecological system, the reduction of greenhouse gas emissions can be taken as the environmental factor of interest. Policy-makers can design and implement  economic incentives aiming at adapting individual decisions to collectively agreed goals  \cite{sullivan2017corporate}. Hence, the incentive-based policy proposed in this work shows the power of incentive-based control policy and leads to some directions in mitigating the global warming issue.

In this paper, we have used a more general and realistic model, replicator-mutator dynamics, to describe
the strategic evolution, but the environment dynamics is still described by the simplified logistic model. Thus, it is of great interest to study the game dynamics under different types of feedback of the practical environmental resource, such as renewing or decaying resource.
The replicator or replicator-mutator equations describe the evolutionary game dynamics in an infinite well-mixed population. Due to this shortcoming, they are not suitable to study game dynamics in  finite or structured populations. Therefore, another direction of future studies is to develop more appropriate individual-based models to study the game dynamics under the environmental feedback. Last but not least, in addition to the incentive-based control, there could exist other kinds of control policies that can be more  effective in various specific situations.

\bibliographystyle{plain} 
\bibliography{main}             

\appendix

\section{Proof of Lemma \ref{alphabetagammadelta}}
\begin{pf}
The equations (\ref{eqn:linearequations}) can be taken as a set of linear equations with respect to unknown variables $\alpha$, $\beta$, $\gamma$, and $\delta$.
It suffices to show that (\ref{eqn:linearequations}) always has solutions. This can be done by checking the ranks of the coefficient matrix $C$  and augmented matrix $[C|D]$, which are given by
\begin{equation}
\begin{bmatrix}
&\begin{aligned}
&(-c+d\\
&~-a+b)
\end{aligned}&\begin{aligned}&(-c+d\\&-a+b)\end{aligned}&0&0\\
&(a-b)&(a-b)&0&0\\
&\begin{aligned}&(-c+2d\\&-a+2b)\end{aligned}&(d+b)&-(\theta+1)&-(\theta+1)\\
&(a-2b)&-b&(\theta+1)&0\\
&-(d+b)&0&1&1\\
\end{bmatrix},  
\end{equation}
\begin{equation}
\begin{bmatrix}
&\begin{aligned}
&(-c+d\\
&~-a+b)
\end{aligned}&\begin{aligned}&(-c+d\\&-a+b)\end{aligned}&0&0&\begin{aligned}&3(c-d\\
&+a-b)\end{aligned}\\
&(a-b)&(a-b)&0&0&3(b-a)\\
&\begin{aligned}&(-c+2d\\&-a+2b)\end{aligned}&(d+b)&-(\theta+1)&-(\theta+1)&\varpi\\
&(a-2b)&-b&(\theta+1)&0&\begin{aligned}&(4b-2a\\
&-\theta-1)\end{aligned}\\
&-(d+b)&0&1&1&b+d-2\\
\end{bmatrix},
\end{equation}
where $\varpi=2(c-2d+a
-2b+\theta+1)$.

One can easily check that $\text{rank} ~C \equiv \text{rank} ~[C|D] \leq 4$, which implies that (\ref{eqn:linearequations}) has at least one set of solutions.
Actually when $-c+d-a+b=0$ and $a-b=0$, (\ref{eqn:linearequations}) will have infinite many solutions; otherwise, it will have exactly one solution.
\hfill$\qed$
\end{pf}

\section{Proof of Lemma \ref{nolimitcycle}}\label{proofoflemma3}
\begin{pf}
For the boundary of $\mathcal{I}$, the same argument presented in the proof of Lemma \ref{nonclosedorbit} is also  applicable to exclude  the existence of periodic orbits that lie on or intersect with the boundary. Thus we only need to focus on the case of the interior $\text{int}(\mathcal{I})$.

When $\mu=0$, the divergence of $\varphi f(x,r)$ is
\begin{equation}\label{eqn:div=0}
    \frac{\partial (\varphi f_1)}{\partial x}(x,r)+\frac{\partial (\varphi f_2)}{\partial r}(x,r)=\varphi(x,r)[(1+\alpha) b-1-\gamma].
\end{equation}
If there exist  periodic orbits in $\text{int}(\mathcal{I})$, 
the Bendixson-Dulac criterion would imply that either $\varphi(x,r)[(1+\alpha) b-1-\gamma]$ is identically zero or it changes sign. Because $\varphi(x,r)$ is strictly positive, we have $(1+\alpha) b-1-\gamma=0$. Then (\ref{eqn:div=0}) turns out to be
\begin{equation}\label{eqn:div=01}
 \frac{\partial (\varphi f_1)}{\partial x}(x,r)=-\frac{\partial (\varphi f_2)}{\partial r}(x,r).  
\end{equation}
Thus system (\ref{eqn:closeloop}) has a \emph{first integral} [\cite{Singer:92}] in $\text{int}(\mathcal{I})$ 
\begin{equation}
    V(x,r)=\int \varphi f_2 dx- \varphi f_1 dr,
\end{equation}
whose derivative over time $t$ satisfies
\begin{equation*}
 \frac{dV}{dt}=\frac{\partial V}{\partial x}\dot{x}+\frac{\partial V}{\partial r}\dot{r}=\varphi f_1f_2-\varphi f_1f_2\equiv 0,
\end{equation*}
and hence $V(x,r)$ remains constant along the solutions of (\ref{eqn:closeloop}).

By definition, a limit cycle is an isolated periodic orbit, which is the $\alpha$-limit set or $\omega$-limit set for some initial points that do not lie on the orbit [\cite{Hirsch:04}]. Suppose there is a limit cycle $\Gamma_0$ in $\text{int}(\mathcal{I})$, and an arbitrary trajectory starts from the initial point $(x_0, r_0)$ and spirals toward $\Gamma_0$ as $t\rightarrow +\infty$ or $t\rightarrow -\infty$. Then according to  \cite[Corollary 10.1]{Hirsch:04}, $(x_0, r_0)$ must have a neighborhood $\Omega$ such that $\Gamma_0$ is the $\omega$-limit set for all points in it. Due to the fact that $V(x,r)$ is the first integral for system (\ref{eqn:closeloop}), $V(x,r)$ must be constant in $\Omega$ because of the continuity. However, the partial derivatives of $V(x,r)$
\begin{equation*}
\begin{aligned}
  &\frac{\partial V}{\partial x}(x,r)=2\varphi(x,r) f_2(x,r)\\
  &\frac{\partial V}{\partial r}(x,r)=-2\varphi(x,r) f_1(x,r).
\end{aligned}  
\end{equation*}
only equal $0$ at $(x^*,r^*)$, which  leads  to  a contradiction. Therefore, system (4) cannot have limit cycles in  $\mathcal{I}$ when $\mu= 0$.
\hfill$\qed$
\end{pf}

\section{Computation of first Lyapunov coefficient}\label{Lyapunovcoefficient}
The 
following is the process of calculating $\ell_1$ as presented in \cite{Kuznetsov:04}. 
Consider a planar autonomous system $\dot{y}=f(y,\mu)$ where $y\in \mathbb{R}^2$ and $\mu\in \mathbb{R}$. Let $A_0$ be the Jacobian matrix evaluated at the equilibrium $y_0$ and the Hopf bifurcation point $\mu_h$. Suppose $A_0$ has two purely imaginary
complex conjugate eigenvalues, given by $\pm i\omega_0$. The algorithm for computing $\ell_1$ can be summarized as follows:
\begin{enumerate}
    \item compute the complex vectors $q, p\in \mathbb{C}^2$ such that 
    \[A_0q=i\omega_0q,~ ~A_0^Tp=-i\omega_0p,~ \langle p,q \rangle=1\]
    where $\langle p,q \rangle=\bar{p}^Tq$ is the inner product;
    \item compose the expression with a complex variable $z$
    \[H(z,\bar{z})=\langle p,f(y_0+zq+\bar{z}\bar{q},\mu_h) \rangle,\]
    and find the coefficients $g_{20}$, $g_{11}$, and $g_{21}$ in the formal Taylor series
    \[H(z,\bar{z})=i\omega_0z+\sum_{j+k\geq 2}\frac{1}{j!k!}g_{jk}z^j\bar{z}^k;\]
    \item the first Lyapunov coefficient $\ell_1(y_0,\mu_h)$ is given by 
\begin{equation*}\label{lcformula}
   \ell_1(y_0,\mu_h)=\frac{1}{2\omega_0^2} \Re (ig_{20}g_{11}+\omega_0g_{21}).
\end{equation*}
\end{enumerate}
Consider system (\ref{eqn:closeloop}). When $\mu=\mu_1$, we have 
\begin{equation}
A_0=\begin{bmatrix}
0&\frac{-\theta\zeta}{(\theta+1)^3}\\
\frac{(\theta+1)(c\theta+d\theta^2-\mu_1\hat{\theta})(a\theta+b\theta^2+\mu_1\hat{\theta})}{\theta^2\zeta^2}&0
\end{bmatrix},
\end{equation}
\begin{equation}
\omega_0=\sqrt{\frac{(c\theta+d\theta^2-\mu_1\hat{\theta})(a\theta+b\theta^2+\mu_1\hat{\theta})}{\theta(\theta+1)^2\zeta}}.   
\end{equation}
Then following the above process, we obtain 
\begin{equation}\label{firstlc}
\begin{aligned}
\ell_1(\mu_1)&=\\
&\frac{\theta(bc-ad)\zeta((b+d)(\theta^3+2\theta)+(a+c)(2\theta^3+1))}{2\omega_0^3(\theta+1)^4\Delta}.
\end{aligned}
\end{equation}

\section{Proof of Lemma \ref{lemma9} }\label{boundaryequilibrium}
\begin{pf}
One only needs to check the case of $\mathcal{B}_b$, since the other case is analogous. When $r(0)=0$ initially, we have $\dot{r}=0$, and thus one can focus on the following one-dimensional dynamics about $x$:
\begin{equation} \label{eq:onedimension}
    \dot{x}=x(1-x)[ax +b(1-x)]+\mu(1-2x), ~~x\in[0,1].
\end{equation}
Thus, the equilibrium's $x$-coordinates are the solutions of the following equation
\begin{equation}\label{equilibriabd}
 x(1-x)[ax+b(1-x)]=\mu(2x-1), ~~x\in[0,1].   
\end{equation}
The left hand side (LHS) of (\ref{equilibriabd}) is non-negative in the interval $x\in [0,1]$ and is exactly equal to $0$ at $x=0,1$, while the right hand side (RHS) of (\ref{equilibriabd}), $\mu(2x-1)$, is negative for $x\in [0,1/2)$ and positive for $x\in (1/2,1]$ when $\mu\in (0,1]$.  Thus, the LHS and RHS of (\ref{equilibriabd}) will be equal at some $x\in (1/2,1)$.

Now, we will show that the LHS and RHS of (\ref{equilibriabd}) will equal for exactly one value of $x$ in the range at $ (1/2, 1)$.
Denote the LHS of (\ref{equilibriabd}) by $\phi(x)$. Its first derivative is $\phi'(x)=-3(a-b)x^2+2(a-2b)x+b$,  and the second derivative is $\phi''(x)=2(a-2b)-6(a-b)x$. We now discuss in the following cases:
\begin{enumerate}[1).]
    \item When $a-b=0$, one has $\phi''(x)=-2b<0$ as $b>0$, which implies $\phi(x)$  is concave. In this case, $\phi(x)$ equals the RHS of (\ref{equilibriabd}) exactly once when $x \in (1/2, 1)$.
    \item When $a-b\neq 0$, the solution to $\phi''(x)=0$ is $x=\frac{a-2b}{3(a-b)}$. Then, 
    \begin{enumerate}[i.]
        \item when $\frac{a-2b}{3(a-b)}\leq 0$  or $\frac{a-2b}{3(a-b)}\geq 1$, one has $b<a\leq 2b $ or $a<b \leq 2a $, respectively. In both cases,   $\phi''(x)$ is non-positive for $x\in [0,1]$. Thus, $\phi(x)$ is concave in $[0,1]$, which results in the same conclusion as in case 1).
        \item when $0<\frac{a-2b}{3(a-b)}< 1$, one has $a>2b$ or $b>2a$. If $a>2b$, we have $\frac{a-2b}{3(a-b)}<\frac{1}{3}$, which results in $\phi''(x)<0$ for $x\in (1/3,1]$. Namely, $\phi(x)$ is concave in $x\in (1/3,1]$. The same conclusion follows. In the other case,
        if $b>2a$, we have $\frac{a-2b}{3(a-b)}>\frac{1}{2}$. Then, one can check that $\phi'(x)$ is negative for $x\in (1/2,1]$. The strictly decreasing function $\phi(x)$ will only be  equal to the RHS of (\ref{equilibriabd}) for one time when $x\in (1/2,1]$.
    \end{enumerate}
\end{enumerate}

When $\theta\geq 1$, the latter statement of the lemma holds trivially since $1/(\theta+1)\leq 1/2$.
For the case $\theta<1$, one can substitute $x=1/(\theta+1)$ into the RHS of (\ref{eq:onedimension}), which yields $\frac{\mu(\theta-1)(\theta+1)^2+\theta(a+\theta b)}{(\theta+1)^3}$. This value  is positive for any $\mu \in(0,1]$ because of (\ref{condition1}). Since the RHS of (\ref{eq:onedimension}) is negative at $x=1$, then due to continuity, the RHS of (\ref{eq:onedimension}) equals $0$  when $x \in (1/(\theta+1), 1)$.
\hfill$\qed$
\end{pf}

\section{Proof of Lemma \ref{lem:repelling}}\label{proof:repelling}
\begin{pf}
The proof is inspired by  \cite[Theorem 12.2.1]{Hofbauer:98}.  As stated before,  the vectors on  the left side $\mathcal{B}_l$ and the right side $\mathcal{B}_r$ point inwards to $\mathrm{int} ( \mathcal{I})$. This implies that these two sides are repelling. Therefore, the subsequent discussion is restricted to the top and bottom sides $\mathcal{B}_t$ and $\mathcal{B}_b$, which are positively invariant.

We define a partial distance function $F:[0,1] \to \mathbb{R}$ as below:
\begin{equation}
    F(r)=r(1-r),
\end{equation}
which indicates the distance of the system states $(x,r)$ to the top and bottom sides $\mathcal{B}_t$ and $\mathcal{B}_b$. It is positive in $\text{int} (\mathcal{I})$ and only equals zero on these two sides. In addition, as $r$ increases from $0$ to $1$, the function value $F(r)$ monotonically increases from $0$ to the maximal value at $r=1/2$ and then monotonically decreases to $0$ at $r=1$. The derivative of $F$ with respect to time $t$ is 
\begin{equation}
\begin{aligned}
\dot{F}(r)=\dot{r}(1-2r)= F(r)\Phi(x,r),
\end{aligned}
\end{equation}
where $\Phi(x,r):= (1-2r)[(\theta+1)x-1]$. We want to show that there exists a $\hat{\delta}>0$ such that $\liminf_{t\to\infty} F(r(t)) > \hat{\delta}$, which implies that $\mathcal{B}_t$ and $\mathcal{B}_b$ are repelling. We only discuss the case of  $\mathcal{B}_b$ here, as the analysis for  $\mathcal{B}_t$ is similar and thus omitted. Denote the unique interior equilibrium by $q^*=(x^*, r^*)$, as given before. Choose $0<r'<\min\{r^*, 1/2\}$, and let $m=F(r')=r'(1-r')$. We define a lower region by $I_l(m) := \{(x,r) \in \mathcal{I} : 0 < F(r) \le m, 0 < r \le r'\}$. Note that $I_l(m) \cap \mathcal{B}_b = \emptyset$, and the condition $0 \le F(r) \le m$ is redundant in this definition, but it simply indicates that the set is related to the  partial distance function evaluated at $r=r'$. Denote the solution of the dynamics \eqref{eqn:closeloop} by $\xi(t):=(x(t), r(t))$, given some initial condition $\xi(0)=(x(0), r(0)) \in \mathcal{I}$. 

First we show that if $\xi(0) \in I_l(m)$, then there exists a time instant $T>0$, such that $\xi(T) \notin I_l(m)$. Define a line $L_c$ in the lower region $I_l(m)$ by $L_c:=\{(x,r) \in I_l(m): x=1/(\theta+1)\}$. Note that on this line, $\dot{r}=0$ and $\dot{x}\ge \epsilon>0$ for some positive constant $\epsilon$, and thus the vectors point horizontally towards the positive $x$-direction. The line $L_c$ divides the lower region $I_l(m)$ into the left region $J_L:=\{(x,r) \in I_l(m): 0 \le x <1/(\theta+1)\}$ and the right region $J_R:=\{(x,r) \in I_l(m) : 1/(\theta+1)< x \le 1\}$. Note that $\dot{F}=F(r)\Phi(x,r)>0$ in $J_R$ while $\dot{F}<0$ in $J_L$. Therefore, if $\xi(0) \in J_R$, the trajectory of $r(t)$ will increase, and there exists a $T>0$ such that $\xi(T) \notin I_l(m)$ (due to the directions of vectors on the line $L_c$, the right side $\mathcal{B}_r$ and $\dot{F}(r) \ge \epsilon'>0$ for some positive constant $\epsilon'$ in a sufficiently large compact set $\Omega$ satisfying $\{(x,r) \in J_R : r = r'\} \subset \Omega \subset J_R$ and $\xi(0) \in \Omega$). If $\xi(0) \in J_L$, then the trajectory $\xi(t)$ will not always stay in $J_L$ due to the Poincar{\'e}-Bendixson theorem, since there are no equilibria in $J_L$. As $\dot{F}(r)<0$ in $J_L$, the trajectory of $r(t)$ will decrease, and $x(t)$ will cross the center line $L_c$, and enter the right region $J_R$. Due to the uniqueness of solutions of \eqref{eqn:closeloop}, similar to the previous analysis of the right region, there exits a $T>0$ such that $\xi(T) \notin I_l(m)$. 

Define $\overline{I_l}(m):=I_l(m) \cup \mathcal{B}_b$. Next we show that there exists $n \in (0,m)$, such that if $\xi(0) \notin \overline{I_l}(m)$, then $\xi(t) \notin I_l(n) := \{(x,r) \in \mathcal{I}: 0 < F(r) \le n, 0 < r \le r'\}$ for all $t \ge 0$. For $z \in \overline{I_l}(m)$, let $T_m(z) := \inf\{T \ge 0 : \xi(0)=z, \xi(T) \notin I_l(m)\}$. Note that when $z \in \mathcal{B}_b$, then $T_m(z)=0$ trivially since $I_l(m) \cap \mathcal{B}_b = \emptyset$. As we have shown previously, any trajectory starting in the lower region $I_l(m)$ will leave this region at some finite time instant. Therefore, we can define $\overline{T}:=\sup_{z \in \overline{T_l}(m)} T_m(z) < \infty$. Let $t_0$ be the first time instant when the trajectory $\xi(t)$ reaches $\overline{I_l}(m)$ (i.e., $t_0 = \min\{T>0 : \xi(T) \in \overline{I_l}(m)\}$), and let $\xi(t_0)=(x(t_0), r(t_0))$. It is obvious that $r(t_0)=r'$ and $F(r(t_0))=m$, namely, the trajectory reaches the boundary of the lower region $I_l(m)$ at time $t_0$. Let $\Phi_{\rm min} = \min_{(x,r) \in \mathcal{I}} \Phi(x,r)$ (the minimum is attainable on the compact set $\mathcal{I}$). Then for $t \in (t_0, t_0+\overline{T})$, we have
\begin{equation} \label{eq33}
    \frac{1}{t-t_0} \int_{t_0}^{t} \Phi(x,r) dt \ge \Phi_{\rm min}.
\end{equation}
Since $\Phi(x,r) = \dot{F}(r) / F(r)$,  the equation \eqref{eq33} can be further derived:
\[
    \frac{1}{t-t_0} \int_{t_0}^{t} \frac{d}{dt} \big( \ln F(r) \big) dt = \frac{1}{t-t_0} \ln \frac{F(r(t))}{F(r(t_0))} \ge \Phi_{\rm min},
\]
which implies
\[
    F(r(t)) \ge F(r(t_0)) e^{\Phi_{\rm min}(t-t_0)} > m e^{\Phi_{\rm min} \overline{T}}.
\]
Let $n=m e^{\Phi_{\rm min} \overline{T}}$. Therefore we have shown that the trajectory will not reach $I_l(n)$ for $t \in (t_0, t_0+\overline{T})$. Since $\overline{T}$ is the supremum of the duration when the solution could stay in the lower region $I_l(m)$, then there exists $T'' \in (t_0, t_0 + \overline{T}]$ such that $\xi(T'') \notin I_l(m)$. Therefore, the trajectory $\xi(t)$ will leave the lower region $I_l(m)$ without reaching $I_l(n) \subset I_l(m)$. Repeating this argument, we conclude that the trajectory $\xi(t)$ will never reach $I_l(n)$. Thus we have shown that there exists $0<\hat{\delta}<n$ such that $\liminf_{t\to\infty} F(r(t)) > \hat{\delta}$, and thus the lower side $\mathcal{B}_b$ is repelling.
\hfill$\qed$
\end{pf}

\section{Functions Expressions in Proof of Lemma \ref{lem:limitcycle}}\label{detailedexpressions}
\begin{enumerate}[(1)]
    \item One does not need to obtain the explicit expression of $G(\tilde{x})$,  since the integrand, $g(s)=\frac{2s}{\alpha(s)}=\frac{2s}{1/4-s^2}$ is an odd function. Obviously, we  have $G(\nu)=\int_0^{\nu}g(s)ds=-\int^0_{-\nu}g(s)ds=G(-\nu)$;
    \item $\dfrac{F(\tilde{x},\tilde{r})}{\varphi(\tilde{r})}=\dfrac{2\mu\tilde{x}-8\mu_1\tilde{x}(1/4-\tilde{x}^2)}{(1/4-\tilde{x}^2)(b+d)\tilde{r}}$;
    \item $g(\tilde{x})F(\tilde{x},\tilde{r})=\dfrac{4\mu\tilde{x}^2-16\mu_1\tilde{x}^2(1/4-\tilde{x}^2)}{{(1/4-\tilde{x}^2)}^2(\tilde{r}+r'^*)(1-\tilde{r}-r'^*)}$;
    \item $F(\tilde{x},\tilde{r})=\dfrac{2\mu \tilde{x}-8\mu_1\tilde{x}(1/4-\tilde{x}^2))}{(1/4-\tilde{x}^2)(\tilde{r}+r'^*)(1-\tilde{r}-r'^*)}$.
\end{enumerate}

\section{Proof of Lemma \ref{equilibriumbu}}\label{equilibriumxustar}
\begin{pf}
In view of the system equations (\ref{eqn:closeloopwithcontrol}),
the $x$-coordinates of the equilibria on $\mathcal{B}_t$ are obtained by solving the following polynomial equation
\begin{equation}\label{equilibriumonupperside}
x(1-x)[x(d-c)-d+u]=\mu(2x-1), ~~x\in[0,1].  
\end{equation}
The RHS of (\ref{equilibriumonupperside}), $\mu(2x-1)$, is a linear function of $x$ which equals $0$ at $x=1/2$. It is negative for $x\in (0,1/2)$ and positive for $x\in (1/2,1)$ when $\mu\in (0,1]$.
The LHS of (\ref{equilibriumonupperside}) equals $0$ at the two end points $x=0,1$. 

When $u>(c+d)/2$ and $c=d$, the LHS will be always positive for $x\in(0,1)$, such that the root of (\ref{equilibriumonupperside}) is in the range $(1/2,1)$. The value of LHS will change the sign from positive to negative (resp. negative to positive) as $x$ increases from $0$ to $1$ when $c>d$ (resp. $c<d$), and it equals $0$ also at $x=(d-u)/(d-c)$. 
One has $(d-u)/(d-c)<1/2$ and $(d-u)/(d-c)>1/2$ respectively for $c<d$ and $c>d$. 
For both cases, there will be some $x\in(1/2,1)$ such that the RHS and LHS are equal as shown geometrically in Fig. (\ref{fig:curveintersection}).
Therefore, one can conclude that there is always one equilibrium with  $x$-coordinate being $1/2<x_t^*<1$.

\begin{figure}[htbp!]
    \centering
\includegraphics[width=6cm]{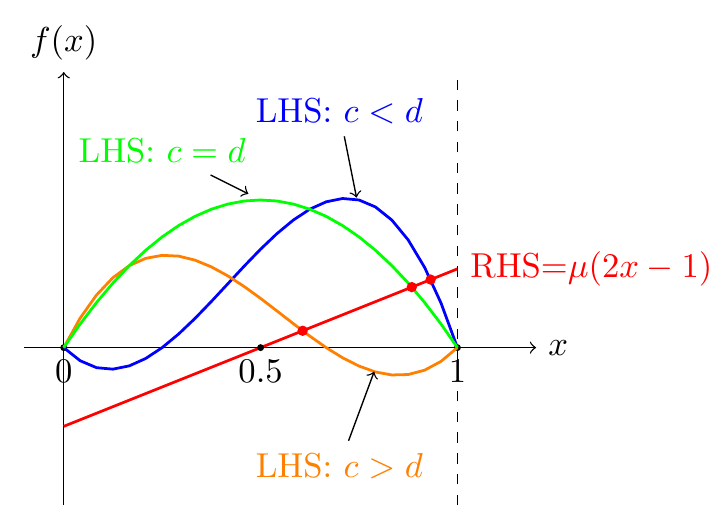}
    \caption{The plot of the LHS and RHS of (\ref{equilibriumonupperside}) as functions of $x$. In all cases, the curves of LHS and RHS have exactly one intersection (red dot) in the range $x\in(1/2,1)$. }
    \label{fig:curveintersection}
\end{figure}
Next, only in the case $c<d$, the curves of the LHS and RHS can have one or two intersections in the range $0<x<1/2$. Thus, the other possible equilibria on $\mathcal{B}_t$ can only be located in $x\in(0,1/2)$.
\hfill$\qed$
\end{pf}

\end{document}